\documentclass{article}

\usepackage{microtype}
\usepackage{graphicx}
\usepackage{booktabs} 

\usepackage{hyperref}

\usepackage[arxiv]{icml2020}

\usepackage[T1]{fontenc}
\usepackage[utf8]{inputenc}
\DeclareUnicodeCharacter{200E}{}
\PassOptionsToPackage{table}{xcolor}

\usepackage{amsmath,amsfonts,bm}

\def\eqref#1{equation~\ref{#1}}

\def\1{\bm{1}}

\def\vc{{\bm{c}}}

\def\vg{{\bm{g}}}

\def\vu{{\bm{u}}}

\def\vx{{\bm{x}}}

\def\evc{{c}}

\DeclareMathAlphabet{\mathsfit}{\encodingdefault}{\sfdefault}{m}{sl}
\SetMathAlphabet{\mathsfit}{bold}{\encodingdefault}{\sfdefault}{bx}{n}

\usepackage{caption}

\usepackage{overpic}
\usepackage{wrapfig}
\usepackage{hyperref}
\usepackage{url}
\usepackage{graphicx}
\usepackage{subcaption}
\usepackage[normalem]{ulem}

\usepackage[table]{xcolor}

\usepackage{booktabs}

\rowcolors{2}{gray!20}{white}
\newcommand\tabularwidth{0.7}

\newcommand{\todo}[1]{}

\newcommand\hide[1]{}

\newcommand\eg{e.g.,~}

\newcommand{\myreffig}[1]{Figure~\ref{fig:#1}}
\newcommand{\myrefeq}[1]{Equation~\ref{eq:#1}}
\newcommand{\myrefsec}[1]{Section~\ref{sec:#1}}
\newcommand{\myreftab}[1]{Table~\ref{tab:#1}}
\newcommand{\myrefalg}[1]{Algorithm~\ref{alg:#1}}

\icmltitlerunning{Latent Space Subdivision}

\begin{document}

\twocolumn[
\icmltitle{Latent Space Subdivision: \ \\ Stable and Controllable Time Predictions for Fluid Flow}



\icmlsetsymbol{equal}{*}

\begin{icmlauthorlist}
\icmlauthor{Steffen Wiewel}{tum}
\icmlauthor{Byungsoo Kim}{eth}
\icmlauthor{Vinicius C. Azevedo}{eth}
\icmlauthor{Barbara Solenthaler}{eth}
\icmlauthor{Nils Thuerey}{tum}
\end{icmlauthorlist}

\icmlaffiliation{tum}{Department of Computer Science, Technical University of Munich, Germany}
\icmlaffiliation{eth}{Department of Computer Science, ETH Zurich, Switzerland}

\icmlcorrespondingauthor{Steffen Wiewel}{wiewel@in.tum.de}

\icmlkeywords{Machine Learning, ICML, Fluid Simulation, Physics, Prediction, LSTM, Autoencoder, Time Series, Deep Sequence Learning, Fluid Flow, Fluids, Smoke}

\vskip 0.3in
]



\printAffiliationsAndNotice{}  


\begin{abstract}
We propose an end-to-end trained neural network architecture to robustly predict the complex dynamics of fluid flows with high temporal stability.
We focus on single-phase smoke simulations in 2D and 3D based on the incompressible Navier-Stokes (NS) equations, which are relevant for a wide range of practical problems.
To achieve stable predictions for long-term flow sequences, a convolutional neural network (CNN) is trained for spatial compression in combination with a temporal prediction network that consists of stacked Long Short-Term Memory (LSTM) layers.
Our core contribution is a novel latent space subdivision (LSS) to separate the respective input quantities into individual parts of the encoded latent space domain.
This allows to distinctively alter the encoded quantities without interfering with the remaining latent space values and hence maximizes external control.
By selectively overwriting parts of the predicted latent space points, our proposed method is capable to robustly predict long-term sequences of complex physics problems.
In addition, we highlight the benefits of a recurrent training on the latent space creation, which is performed by the spatial compression network.
\end{abstract}


\section{Introduction} \label{sec:intro}

Computing the dynamics of fluids requires solving a set of complex equations over time. This process is computationally very expensive, especially when considering that the stability requirement poses a constraint on the maximal time step size that can be used in a simulation. 

Due to the high computational resources, approaches for machine learning based physics simulations have recently been explored. One of the first approaches used Regression Forest as a regressor to forward the state of a fluid over time~\cite{Ladicky2015}. Handcrafted features have been used, representing the individual terms of the Navier-Stokes equations. These context-based integral features can be evaluated in constant time and robustly forward the state of the system over time. In contrast, using neural networks for the time prediction has the advantage that no features have to be defined manually, and hence these methods have recently gained increased attention. 
In graphics, the presented neural prediction methods~\cite{lsp2019,kim19a,morton2018deep} use a two-step approach, where first the physics fields are translated into a compressed representation, i.e., the latent space. Then, a second network is used to predict the state of the system over time in the latent space. The two networks are trained individually, which is an intuitive approach as spatial and temporal representations can be separated by design. In practice, the first network (i.e., the autoencoder) introduces small errors in the encoding and decoding in each time step. In combination with a temporal prediction network these errors accumulate over time, introducing drifting over prolonged time spans and can even lead to instability, as seen in \myreffig{longterm_prediction} (top right). This is especially problematic in supervised learned latent space representations, since the drift will shift the initial, user-specified conditions (\eg an object's position) into an erroneous latent space configuration originated from different conditions.

Like previous work, we use a neural network to predict the motion of a fluid over time, but with the central goal to increase accuracy and robustness of long-term predictions. 
We propose to use a joint end-to-end training of both components, the fluid state compression and temporal prediction of the motion. A key observation is also the need to control the learned latent space, such that we can influence it to impose boundary conditions and other known information that is external to the simulation state. We therefore propose a latent space subdivision that separates the encoded quantities in the latent space domain. The subdivision is enforced with a latent space split soft-constraint for the input quantities velocity and density, and allows to alter the individual encoded components separately. This separation is a key component to robustly predict long-term sequences.

\section{Related Work and Background} \label{sec:relwork}

Our work concentrates on single-phase flows, which are usually modeled by a pressure-velocity formulation of the incompressible {Navier-Stokes} equations:
\begin{align} \label{eq:navierstokes}
   \frac{\partial \vu} {\partial{t}} + \vu \cdot \nabla \vu &= \frac{-1} {\rho} \nabla{p} + \nu\nabla^2\vu + \vg \\
   \nabla\cdot\vu &= 0, \nonumber
\end{align}
where $p$ denotes pressure, $\vu$ is the flow velocity, and $\rho,\nu,\vg$ denote density, kinematic viscosity, and external forces, respectively.
The density $\rho$ is passively advected 
by the velocity $\vu$.
A complete overview of fluid simulation techniques in computer graphics can be found in~\cite{bridson2015}.

Data-driven flow modelling encompasses two distinguished and complementary efforts: dimensionality reduction and reduced-order modelling.
Dimensionality reduction (e.g., Singular Value Decomposition) scales down the analyzed data into a set of important features in an attempt to increase the sparsity of the representation, while reduced-order modelling (e.g., Galerkin Projection) describes the spatial and temporal dynamics of a system represented by a set of reduced parameters.
In computer graphics, the work of Treuille et al. \yrcite{Treuille2006}~was the first to use Principal Component Analysis (PCA) for dimensionality reduction coupled with a Galerkin Projection method for subspace simulation.
This approach was later extended~\cite{Kim2013} with a cubature approach for enabling Semi-Lagrangian and Mac-Cormack~\cite{Selle:2008:USM} advection schemes, while improving the handling of boundary conditions.
Reduced-order modelling was also explored to accelerate the pressure projection step in liquid simulations \cite{Ando2015}.

Instead of computing reduced representations from pre-simulated velocity fields, alternative basis functions can be used for reduced-order modelling; examples of basis functions include Legendre Polynomials~\cite{Gupta2007}, modular~\cite{Wicke:2009} and spectral~\cite{Long2009} representations.
Also Laplacian Eigenfunctions have been successfully employed for dimensionality reduction, due to their natural sparsity and inherent incompressibility.
De Witt et al. \yrcite{DeWitt2012}~combined Laplacian Eigenfunctions with a Galerkin Projection method, enabling fast and energy-preserving fluid simulations.
The approach was extended to handle arbitrarily-shaped domains~\cite{Liu2015}, combined with a Discrete Cosine Transform (DCT) for compression~\cite{jones2016compressing}, and improved for scalability~\cite{Cui2018}.

The aforementioned methods for data-driven flow modelling use linear basis functions for dimensionality reduction. This enables the use of Galerkin Projection for subspace integration, but it limits the power of the reconstruction when compared to non-linear embeddings. The latent spaces generated by autoencoders (AE) are non-linear and richly capture the input space with fewer variables~\cite{RadfordMC15,wu2016learning}. In light of that, Wiewel et al.~\yrcite{lsp2019} combined a latent space representation with recurrent neural networks (RNN) to predict the temporal evolution of fluid functions in the latent space domain. Kim et al.~\yrcite{kim19a} introduced a generative deep neural network for parameterized fluid simulations that only takes a small set of physical parameters as input to very efficiently synthesize points in the learned parameter space. Their method also proposes an extension to latent space integration by training a fully connected neural network that maps subsequent latent spaces. Our work is related to these two methods, but a main difference is that we use an end-to-end training of both the spatial compression and the temporal prediction. In combination with our latent space subdivision, our predictions are more stable, while previous approaches fail to properly recover long-term integration correspondences due to the lack of autoencoder regularization.

For particle-based fluid simulations, a temporal state prediction using Regression Forest was presented in Ladicky et al.~\yrcite{Ladicky2015}. Handcrafted features are evaluated in particle neighborhoods and serve as input to the regressor, which then predicts the particle velocity of the next time step.
Machine learning has also been used in the context of grid-based (Eulerian) fluid simulations. Tompson et al.~\yrcite{tompson2016accelerating} used a 
convolutional neural network (CNN) to model spatial dependencies in conjunction with an unsupervised loss function formulation to infer pressure fields. A simpler three-layer fully connected neural network for the same goal was likewise proposed \cite{Yang2016}. As an alternative, learned time evolutions for Koopman operators were proposed \cite{morton2018deep}, which however employ a pre-computed dimensionality reduction via PCA. Chu et al.~\yrcite{chu2017cnnpatch} enhance coarse fluid simulations to generate highly detailed turbulent flows. Individual low-resolution fluid patches were tracked and mapped to high-resolution counterparts via learned descriptors. Xie et al.~\yrcite{xie2018tempogan} extended this approach by using a conditional generative adversarial network with a spatio-temporal discriminator supervision. Small-scale splash details in hybrid fluid solvers were targeted with deep learning-based stochastic models~\cite{um2018liquid}. For a review of machine learning applied to fluid mechanics we refer the reader to~\cite{Brunton2019}.

In the context of data-driven flow modelling, methods to interpolate between existing data have been presented. 
Raveendran et al.~\yrcite{Raveendran2014} presented a technique to smoothly blend between pre-computed liquid simulations, which was later extended with more controllability~\cite{Manteaux2016}. Thuerey et al.~\yrcite{Thuerey2016} used dense space-time deformations represented by grid-based signed-distance functions for interpolation of smoke and liquids, and Sato et al.~\yrcite{Sato2018} interpolated velocity fields by minimizing an energy functional.


\section{Method} \label{sec:method}

The central goal of our models is to robustly and accurately predict long-term sequences of flow dynamics. For this, we need an autoencoder to translate high-dimensional physics fields into a compressed representation (latent space) and a temporal prediction network to advance the state of the simulation over time. 
A key observation is that if these two network components are trained individually, neither component has a holistic view on the underlying problem. 
The autoencoder, consisting of an encoder $E$ and a decoder $D$, generates a compressed representation $\vc = E(\vx)$, which focuses solely on the reconstruction $\tilde{\vx} = D(\vc)$ of the given input $\vx$.
Hence, the loss function to minimize is given by \mbox{$\|\vx - \tilde{\vx}\|$}.
Without considering the aspect of time, the autoencoder's latent space only stores spatial descriptors.
Due to the exclusive focus on space, temporally consecutive data points are not necessarily placed close to each other in the latent space domain.
This poses substantial challenges for the temporal prediction network.

\begin{figure}[htb]
   \centering
   \includegraphics[width = 0.3\textwidth]{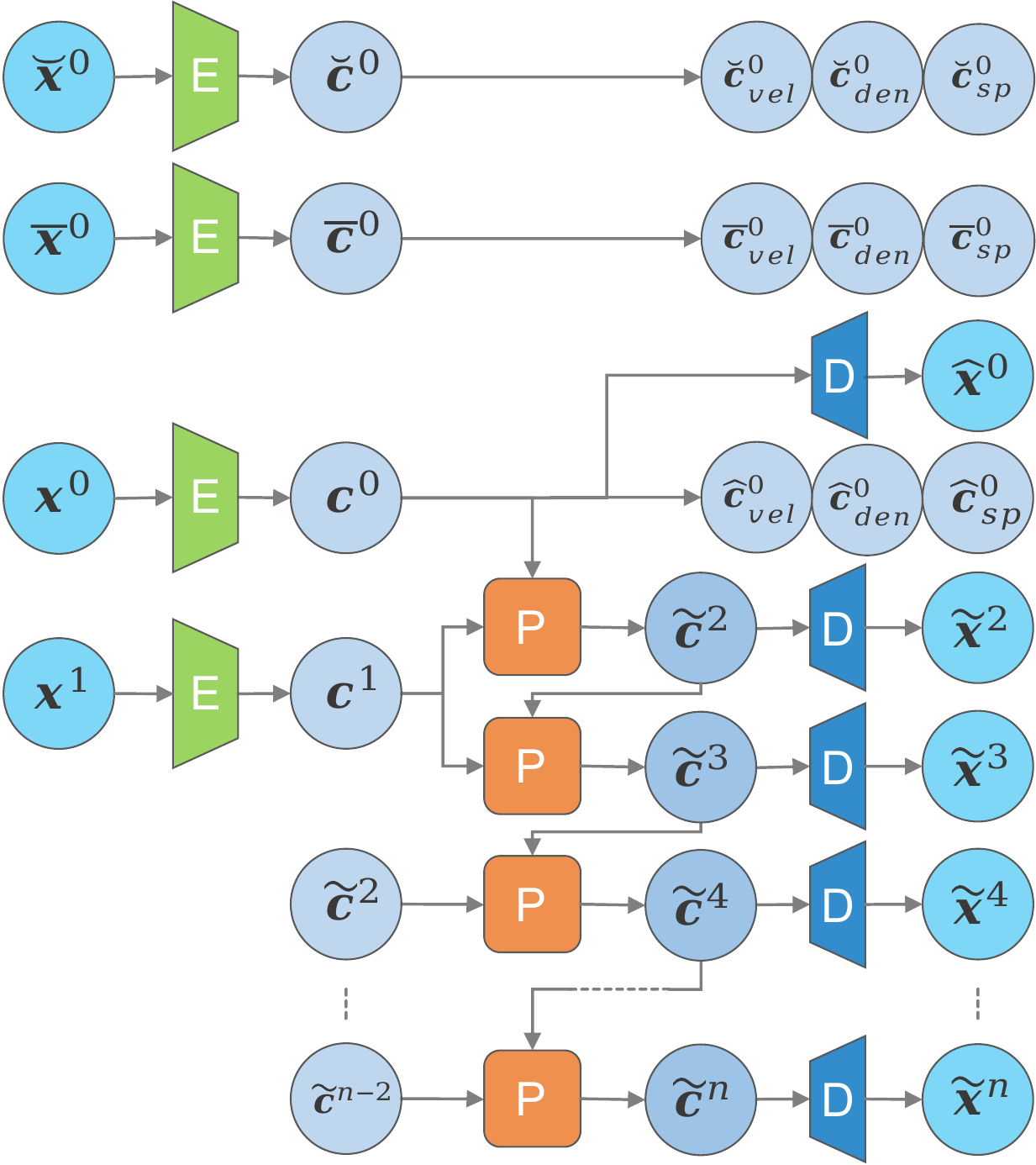}
   \caption{Combined training of autoencoder and temporal prediction. The prediction networks input window is set to $w=2$. Thus, the count of recurrent predictions is $n_i = n-2$. The LSS is enforced by applying the split loss on $\bar{\vc}^0_{den}$ and $\breve{\vc}^0_{vel}$ for velocity and density, respectively. A direct AE reconstruction loss is only performed on $\hat{\vx}^0$. }
   \label{fig:combined_method}
   \vspace{-3.0em}
\end{figure}

Therefore, we consider the aspect of time within the training of the autoencoder in order to shape its latent space with respect to temporal information, in addition to the spatial information.
Thus, we propose an end-to-end training procedure, where we train our autoencoder and temporal prediction network simultaneously by internally connecting the latter as a recurrent block to the encoding and decoding blocks of the spatial autoencoder.
As a result, the latent space domain is aware of temporal changes, and 
can yield temporally coherent latent space points that are suitable for the
time prediction network.
By default, our training process includes the combined training of our spatial autoencoder and temporal prediction network as shown in \myreffig{combined_method}.
In this figure, the encoder E, decoder D and prediction network P are duplicated for visualization purposes.
In the next sections, we describe each individual network in more detail.

\subsection{Spatial Encoding} \label{sec:spatial}
The spatial encoding of the data is performed by a regular autoencoder, the network for which is split into an encoding and decoding part \cite{kim19a}. 
The encoder contains $16$ convolution layers with skip connections, which connect its internal layers, followed by one fully-connected layer.
The decoder consists of a fully-connected layer, which is followed by $17$ convolution layers with skip connections.
For the fluid simulation dataset used in this work, the input is either 2D or 3D, leading to the usage of 2D- or 3D-convolutions and a feature dimension of $3$ or $4$, respectively.
Furthermore, a data-specific curl-layer is appended to the decoder network to enforce zero divergence in the resulting velocity field \cite{kim19a}, as required by the NS equations (see \myrefeq{navierstokes}).

The dimensionality of the latent space $\vc$ for a given $\vx$ is defined by the final layer of the encoder and can be freely chosen.
We pass a velocity $\vu$ as well as a density field $\rho$ to our encoder, i.e., $\vx=[\vu,\rho]$.
The velocity field is an active quantity that is used in fluid simulations to advect a passive quantity forward in time.
In case of smoke simulations, which is a specific instance of fluid simulations, the passive density field is advected by the flow velocity. 
As a result, $\vc$ contains information about both the active and passive fields, i.e., velocity and density.
Hence, we can accurately advect the passive quantity density with a velocity field with low
computational effort, it makes sense to compute the advection outside of the network, and project the new state into the latent space.

In order to be able to alter active and passive fields individually in the compressed representation $\vc$, with given input field $\vx = [\vx_{vel}, \vx_{den}]$ where the subscripts $vel$ and $den$ thereby denote the velocity and density part, we subdivide $\vc$ into separate parts for velocity $\vc_{vel}$ and density $\vc_{den}$, respectively.
This property of the latent space is needed for projecting the new state of the passive quantity into the latent space domain.
Additionally, to exert explicit external control over the prediction, we designate another part of $\vc$ to contain supervised parameters, called $\vc_{sp}$~\cite{kim19a}. 
In our case of, e.g., a smoke simulation with a rotating cup filled with smoke, such supervised parameters can be the position of a smoke source or the rotation angle of a solid obstacle.
With this subdivision, we increase stability of our predictions and allow for explicit external control.
This subdivision is described in \myrefeq{latent_space}, where $v$, $d$, and $sp$ describe the indices of the velocity, density, and supervised parameter parts in the latent space domain, respectively.
\newcommand{\vin}{\scalebox{0.6}[1.5]{\rotatebox[origin=c]{-90}{$\sim$}}}
\begin{align} \label{eq:latent_space}
   \vc &=\big[ \vc_{vel} \, \vin \, \vc_{den} \, | \, \vc_{sp} \big] \\
   \text{where } \quad \vc_{vel} &= \big[ \evc_0, \dots, \evc_v \big], \nonumber \\ 
   \vc_{den} &= \big[ \evc_{v+1}, \dots, \evc_d \big], \nonumber \\ 
   \vc_{sp} &= \big[ \evc_{d+1}, \dots, \evc_{sp} \big]. \nonumber
\end{align}
To arrive at the desired subdivision, the split loss $\mathcal{L}_{split}$ (see \myrefeq{loss_split}) is used as a loss function in the training process.
It is modelled as a soft constraint and thereby does not enforce the parts $\vc_{vel}$ and $\vc_{den}$ to be strictly disjunct. The loss is defined as
\begin{align}
   \mathcal{L}_{split}(\vc, I_s, I_e) &= \sum_{i=I_s}^{I_e} \|\evc_i\|_1.
   \label{eq:loss_split}
\end{align}
Since we divide the latent space in three parts, $\mathcal{L}_{split}$ is applied twice.
For the velocity part $\vc_{vel}$, the indices $I_s = v+1$ and $I_e = d$ are chosen to indicate that the density part must not be used on encoding velocities, i.e., $\mathcal{L}_{split}(\vc, v+1, d)$.
In contrast to the previous limits, for the density part $\vc_{den}$ the velocity part is indicated by choosing the indices $I_s = 0$ and $I_e = v$, i.e. $\mathcal{L}_{split}(\vc, 0, v)$.
First, only the velocity part $\vx_{vel}$ of input $\vx$ is encoded (the density part $\vx_{den}$ is zero, i.e., $\bar{\vx}=[\vx_{vel},0]$), yielding $\bar{\vc}_{vel}$.
Vice versa, the density part $\vx_{den}$ of input $\vx$ is encoded, whereas the velocity part $\vx_{vel}$ is replaced with zeros, i.e., $\breve{\vx}=[0,\vx_{den}]$, resulting in $\breve{\vc}_{den}$.
Therefore, the loss is applied twice for the $\bar{\vc}$ and $\breve{\vc}$ encodings as $\mathcal{L}_{split}(\bar{\vc}, v+1, d)$ and $\mathcal{L}_{split}(\breve{\vc}, 0, v)$, respectively.

In order to exhibit external control over the prediction, $\vc_{sp}$ is enforced to contain parameters describing certain attributes of the simulation.
While training the network, an additional soft-constraint is applied, which forces the encoder to produce the supervised parameters.
The soft-constraint is implemented as the mean-squared error of the values generated by the encoder $\hat\vc_{sp}$ and the ground truth data $\vc_{sp}$
and consitutes the supervised loss $\mathcal{L}_{sup}$ as
\begin{align}
   \mathcal{L}_{sup}( \vc_{sp}, \hat\vc_{sp} ) &= \| \vc_{sp} - \hat\vc_{sp} \|_2^2.
   \label{eq:loss_sup_params}
\end{align}
Additionally, an AE loss $\mathcal{L}_{AE}$ (\myrefeq{loss_AE}) is applied to the decoded field $\hat{\vx}$.
It forces the velocity part of the decoded field to be close to the input velocity by applying the mean-absolute error.
To take the rate of change of the velocites into consideration as well, the mean-absolute error of the velocities' gradient is added to the formulation.
In contrast, the density part is handled by directly applying the mean-squared error on the decoded output density and the input. The AE loss is thereby defined as
\begin{align}
   \mathcal{L}_{AE}(\vx, \hat{\vx}) &= \| \vx_{vel} - \hat{\vx}_{vel} \|_1 \nonumber \\
   &+ \|\nabla \vx_{vel} - \nabla \hat{\vx}_{vel} \|_1 \nonumber \\
   &+ \|\vx_{den} - \hat{\vx}_{den} \|_2^2.
   \label{eq:loss_AE}
\end{align}

\subsection{Time Prediction Network} \label{sec:temporal}

The prediction network performs a temporal transformation of its input to the temporal consecutive state.
The inputs are a series of $w$ consecutive input states.
The prediction network block contains two recurrent LSTM layers, followed by two 1D-convolutional layers. 
In our case of 2D smoke simulations, two consecutive latent space points of dimension $16$ are used as input.
Those are fed to the prediction layers and result in one latent space point of dimension $16$, called the residuum $\Delta\vc^t$.
Afterwards, the residuum is added to the last input state to arrive at the next consecutive state, i.e., $\tilde{\vc}^{t+1} = \vc^t + \Delta\vc^t$.

Due to the subdivision capability of our autoencoder, our temporal prediction network supports external influence over the predictions it generates.
After each prediction, it is possible to replace or update information without the need of re-encoding the predicted data.
Instead, only parts of the predicted latent space point can be replaced, enabling fine-grained control over the flow.
For example, in the case of smoke simulations, the passive smoke density quantity can be overwritten with an externally updated version, 
i.e., the $\vc_{den}$ part is replaced by $\vc$.
This allows for adding new smoke sources or modifying the current flow by removing smoke from certain parts of the simulation domain.

Considering the exposition of the prediction input window $w$, which can be chosen freely, and the desired internal iteration count $n_i = n - w$, the additive prediction error is brought into consideration for the prediction network P while training, i.e., it is traversed $n_i$ times.
This leads to a combined training loss of AE and P defined as
\begin{align}
   \mathcal{L} &= \mathcal{L}_{AE, direct} + \mathcal{L}_{sup} + \mathcal{L}_{split,vel} + \mathcal{L}_{split,den} \nonumber \\
   &+ \sum_{n=0}^{n_i} \big( \mathcal{L}_{AE, pred_{n_i}} \big),
   \label{eq:loss_total}
\end{align}
where $\mathcal{L}_{AE}$ is applied to the corresponding pairs of the decoded outputs $\tilde{\vx}^2 ... \tilde{\vx}^n$ of the prediction network P and their corresponding ground-truth $\vx^2 ... \vx^n$.
Thereby, our final loss is the sum of all the previously presented losses.

In our combined training approach, both networks update their weights by applying holistic gradient evaluations, i.e., are trained end-to-end.
The benefit of the end-to-end training is that the spatial autoencoder AE also incorporates temporal aspects when updating its weights. 
In addition, by recurrently applying $\mathcal{L}_{AE}$ on the predictions, the prediction network P is trained to actively minimize accumulating additive errors.
To incorporate temporal awareness in the autoencoder, the decoder block is connected to the individual prediction outputs and is thereby reused several times in one network traversal.
The recurrent usage of the decoding block is commonly known as weight sharing \cite{Bromley:1993:SVU:2987189.2987282}.
Furthermore, by applying the prediction losses on the decoded predictions, the spatial autoencoder adapts to the changes induced by the temporal prediction as well, which furthers the focus of the autoencoder to produce latent spaces suitable for temporal predictions.
As a result, the prediction network is capable of robustly and accurately predicting long-term sequences of complex fluid flows.


\section{Training Datasets} \label{sec:training}

\begin{figure}[ht]
  \newcommand\subfigwidth{.055\textwidth}
\begin{center}
  \begin{subfigure}{\subfigwidth}
     \centering
     \frame{\includegraphics[width = 0.99\textwidth]{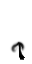}}
  \end{subfigure}%
  \begin{subfigure}{\subfigwidth}
     \centering
     \frame{\includegraphics[width = 0.99\textwidth]{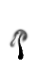}}
   \end{subfigure}%
   \begin{subfigure}{\subfigwidth}
      \centering
      \frame{\includegraphics[width = 0.99\textwidth]{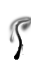}}
   \end{subfigure}%
   \begin{subfigure}{\subfigwidth}
     \centering
     \frame{\includegraphics[width = 0.99\textwidth]{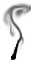}}
   \end{subfigure}%
   \begin{subfigure}{\subfigwidth}
     \centering
     \frame{\includegraphics[width = 0.99\textwidth]{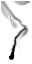}}
   \end{subfigure}%
   \begin{subfigure}{\subfigwidth}
     \centering
     \frame{\includegraphics[width = 0.99\textwidth]{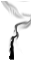}}
   \end{subfigure}%
   \begin{subfigure}{\subfigwidth}
     \centering
     \frame{\includegraphics[width = 0.99\textwidth]{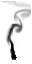}}
   \end{subfigure}%
   \begin{subfigure}{\subfigwidth}
     \centering
     \frame{\includegraphics[width = 0.99\textwidth]{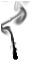}}
   \end{subfigure}%
\end{center}%
\renewcommand\subfigwidth{.055\textwidth}
\begin{center}
  \begin{subfigure}{\subfigwidth}
     \centering
     \frame{\includegraphics[width = 0.99\textwidth]{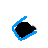}}
  \end{subfigure}%
  \begin{subfigure}{\subfigwidth}
     \centering
     \frame{\includegraphics[width = 0.99\textwidth]{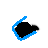}}
   \end{subfigure}%
   \begin{subfigure}{\subfigwidth}
      \centering
      \frame{\includegraphics[width = 0.99\textwidth]{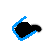}}
   \end{subfigure}%
   \begin{subfigure}{\subfigwidth}
     \centering
     \frame{\includegraphics[width = 0.99\textwidth]{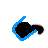}}
   \end{subfigure}%
   \begin{subfigure}{\subfigwidth}
     \centering
     \frame{\includegraphics[width = 0.99\textwidth]{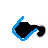}}
   \end{subfigure}%
   \begin{subfigure}{\subfigwidth}
     \centering
     \frame{\includegraphics[width = 0.99\textwidth]{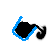}}
   \end{subfigure}%
   \begin{subfigure}{\subfigwidth}
     \centering
     \frame{\includegraphics[width = 0.99\textwidth]{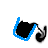}}
   \end{subfigure}%
   \begin{subfigure}{\subfigwidth}
     \centering
     \frame{\includegraphics[width = 0.99\textwidth]{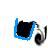}}
   \end{subfigure}%
\end{center}%
\renewcommand\subfigwidth{.055\textwidth}
\begin{center}
  \begin{subfigure}{\subfigwidth}
     \centering
     \frame{\includegraphics[width = 0.99\textwidth]{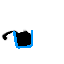}}
  \end{subfigure}%
  \begin{subfigure}{\subfigwidth}
     \centering
     \frame{\includegraphics[width = 0.99\textwidth]{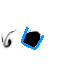}}
   \end{subfigure}%
   \begin{subfigure}{\subfigwidth}
      \centering
      \frame{\includegraphics[width = 0.99\textwidth]{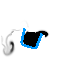}}
   \end{subfigure}%
   \begin{subfigure}{\subfigwidth}
     \centering
     \frame{\includegraphics[width = 0.99\textwidth]{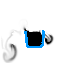}}
   \end{subfigure}%
   \begin{subfigure}{\subfigwidth}
     \centering
     \frame{\includegraphics[width = 0.99\textwidth]{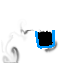}}
   \end{subfigure}%
   \begin{subfigure}{\subfigwidth}
     \centering
     \frame{\includegraphics[width = 0.99\textwidth]{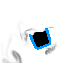}}
   \end{subfigure}%
   \begin{subfigure}{\subfigwidth}
     \centering
     \frame{\includegraphics[width = 0.99\textwidth]{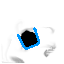}}
   \end{subfigure}%
   \begin{subfigure}{\subfigwidth}
     \centering
     \frame{\includegraphics[width = 0.99\textwidth]{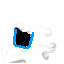}}
   \end{subfigure}%
\end{center}
\caption{ 
Example sequences of our 2D datasets: moving smoke (top), rotating (center) and moving cup (bottom). The smoke density is shown as black and the cup-shaped obstacle in blue.}
\label{fig:mov_smoke_data}
\label{fig:rot_cup_data}
\label{fig:rot_cup_mov_data}
\vspace{-1.0em}
\end{figure}

The datasets we used to train our networks contain randomized smoke flows simulated with an open source framework \cite{mantaflow}.
In total, three different scene setups were used to capture a wide range of complex physical behavior.
The first scene contains a moving smoke inflow source 
that generates hot smoke continuously, which is rising and producing complex swirls (see \myreffig{mov_smoke_data}).

\begin{figure}[b]
  \begin{center}
    \captionsetup{type=table}
    \caption{Statistics of our datasets.
    } \label{tab:training_data}
      \resizebox{\tabularwidth\width}{!}{%
      \begin{tabular}{@{}>{\columncolor{white}[0pt][\tabcolsep]}cccc>{\columncolor{white}[\tabcolsep][0pt]}c@{}}
         \toprule
         \rowcolor{white}
         Scene Type       & Resolution & \# Scene & \# Frames \\
         \midrule
         Rotating and Moving Cup (3D)    & $48^3$  & 100 & 600 \\ \addlinespace
         Moving Smoke (3D)               & $48^3$  & 100 & 600 \\ \addlinespace
         Rotating and Moving Cup (2D)    & $64^2$  & 200 & 600 \\ \addlinespace
         Rotating Cup (2D)               & $48^2$  & 200 & 600 \\ \addlinespace
         Moving Smoke (2D)               & $32x64$ & 200 & 600 \\
         \bottomrule
      \end{tabular}
      }
      \global\rownum=0\relax%
  \end{center}
\end{figure}

The second and third scenes simulate cold smoke in a cup-shaped obstacle. The former rotates the cup randomly around a fixed axis, while the latter additionally applies a translation (see \myreffig{rot_cup_data}).
The rising smoke and the rotating cup scene each expose one control parameter, i.e., movement on the x-axis and rotation around the z-axis, whereas the rotating and moving cup scene exposes both of these control parameters.
Each of the three datasets in 2D contains $200$ randomly generated scenes with $600$ consecutive frames.
Additionally, the moving smoke as well as the rotating and moving cup dataset was generated in 3D with $100$ randomly generated scenes and $600$ consecutive frames (see \myreftab{training_data}). 
\begin{figure*}[hbt]
  \newcommand\subfigwidth{.28\textwidth}
  \centering
  \begin{subfigure}{\subfigwidth}
    \centering
    \includegraphics[width = 1.0\textwidth, trim=60 60 60 60, clip]{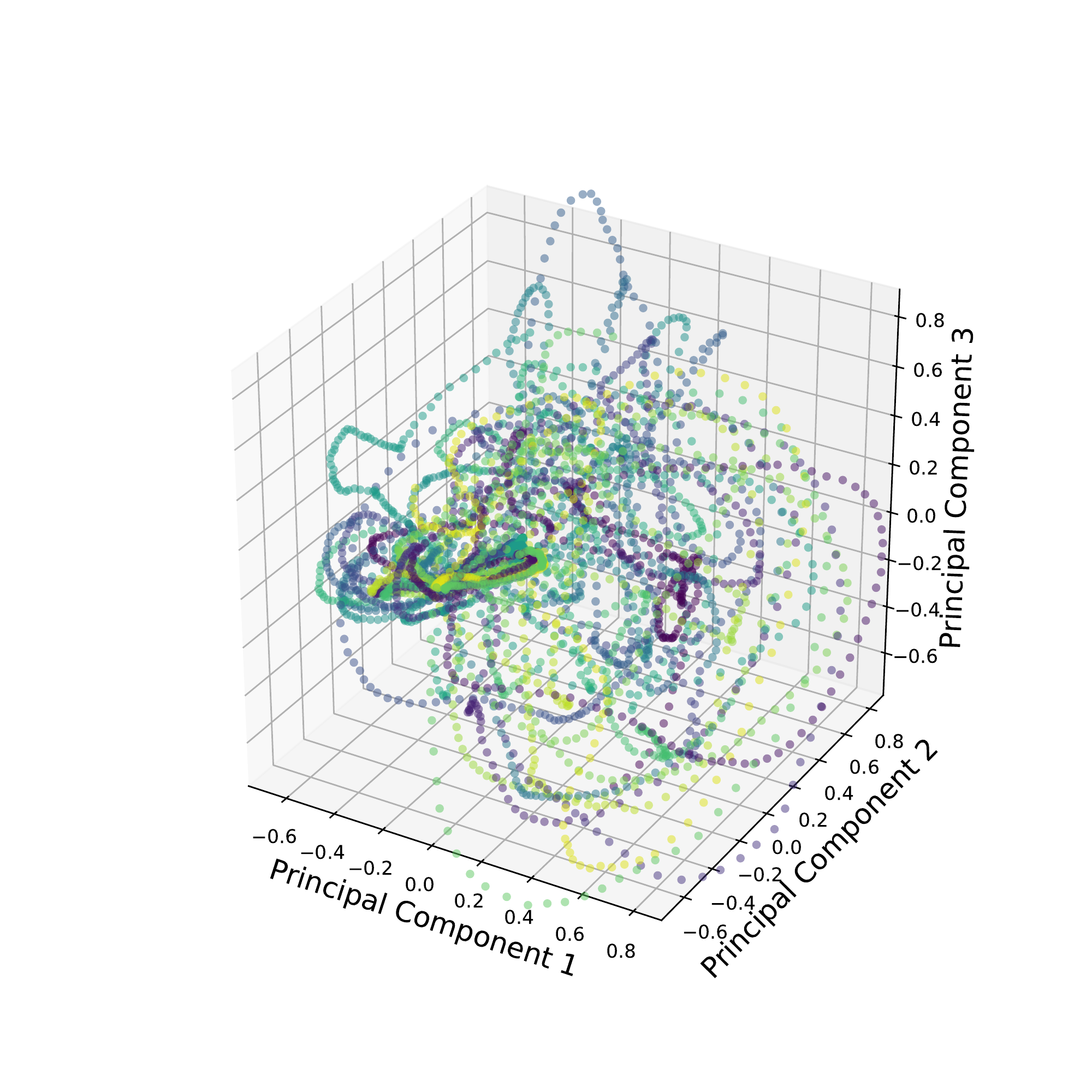}
    \captionsetup{margin=0.03\textwidth}
    \caption{No temporal constraints or supervised parameters.}
    \label{fig:unordered_ls}
  \end{subfigure}%
  \begin{subfigure}{\subfigwidth}
     \centering
     \includegraphics[width = 1.0\textwidth, trim=60 60 60 60, clip]{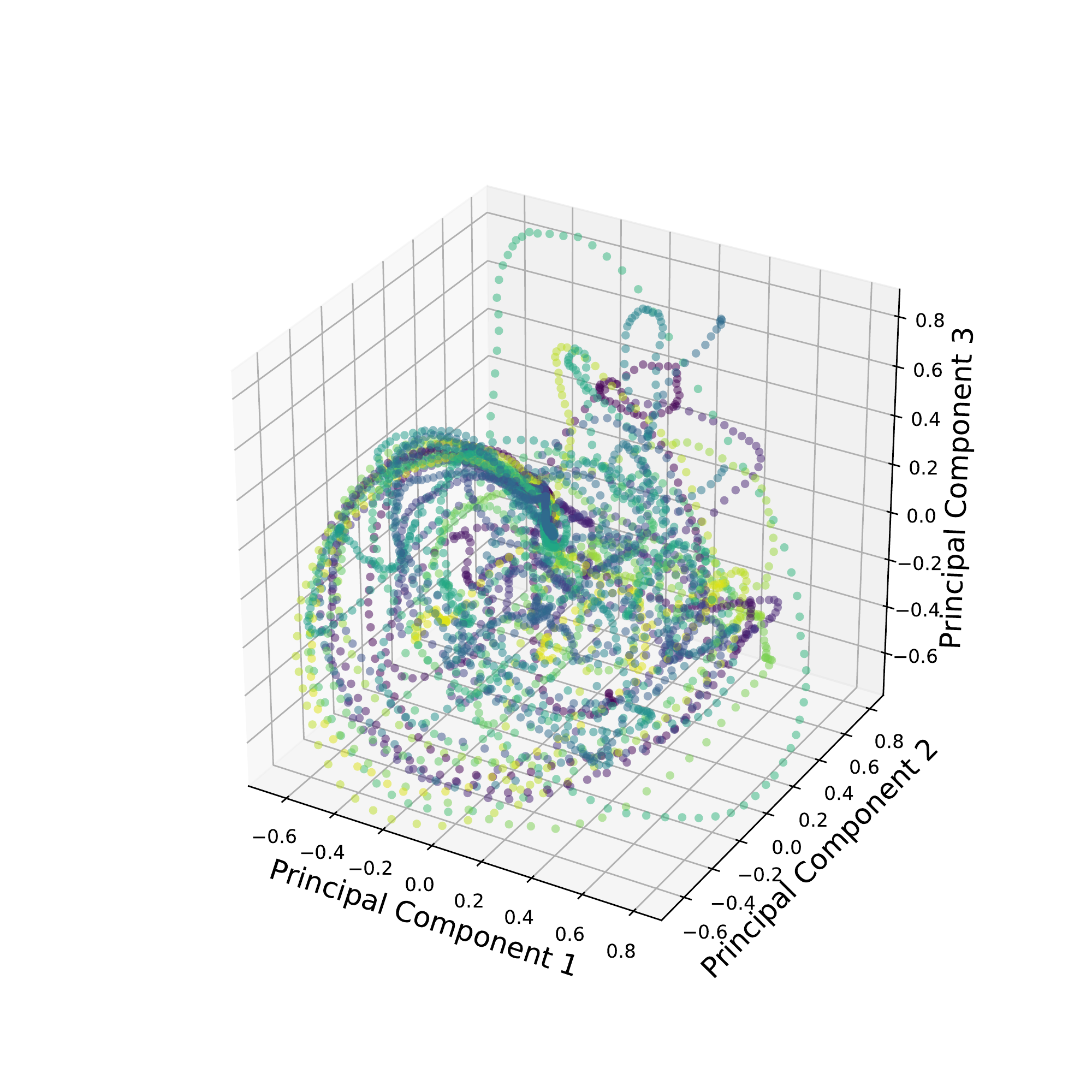}
     \captionsetup{margin=0.03\textwidth}
     \caption{No temporal constraints but with supervised parameters.}
     \label{fig:unordered_ls_sup_param}
   \end{subfigure}%
  \begin{subfigure}{\subfigwidth}
    \centering
    \includegraphics[width = 1.0\textwidth, trim=60 60 60 60, clip]{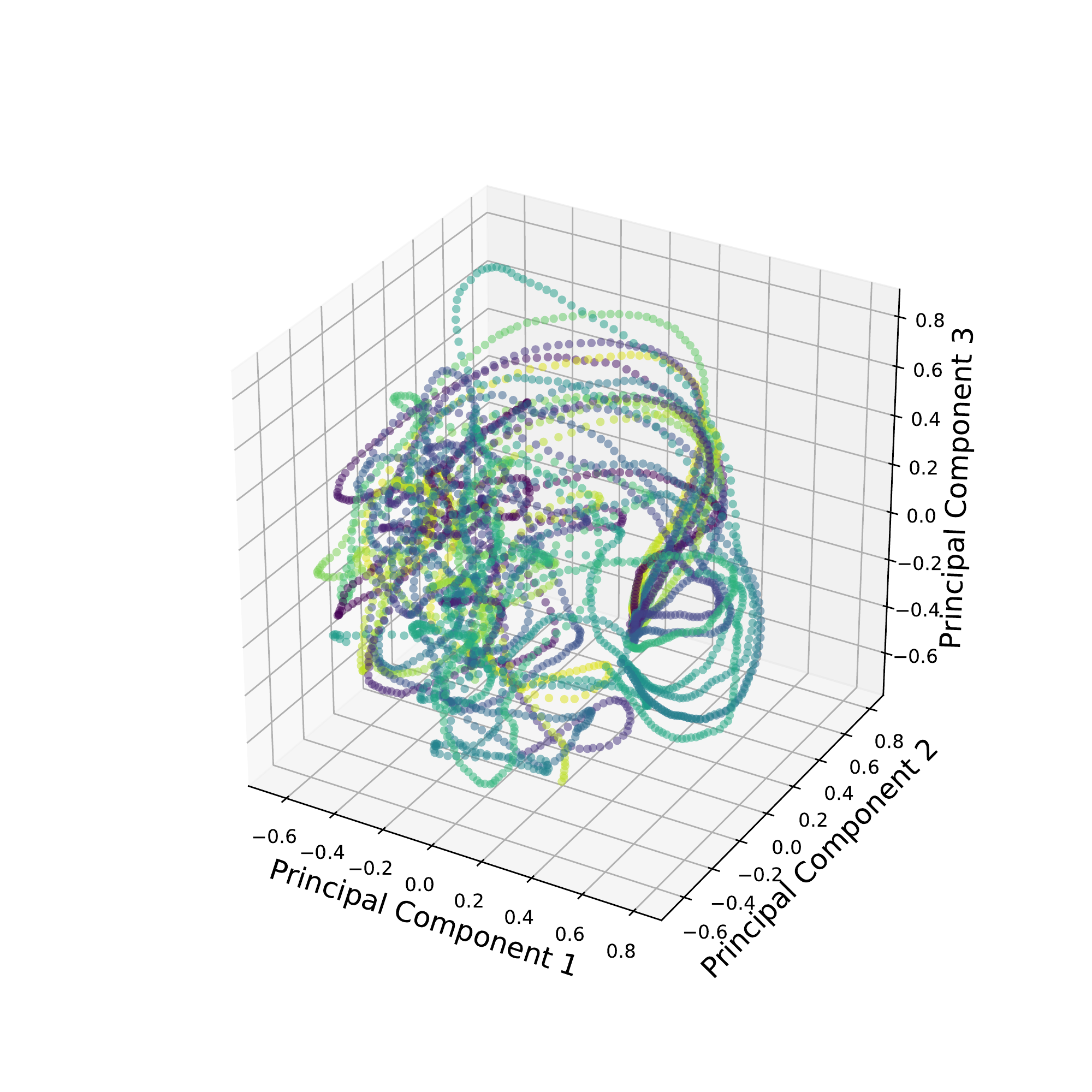}
    \captionsetup{margin=0.03\textwidth}
    \caption{With temporal constraints and supervised parameters (ours).}
    \label{fig:ordered_ls}
  \end{subfigure}%
  \vspace{1em}
  \caption{Spatial encodings of $200$ frames of $20$ different smoke simulations. The latent space points are normalized to their respective maximum and processed with PCA for visualization purposes. Each color stands for a single simulation and represents a series of $200$ frames.}
  \vspace{-1.7em}
  \label{fig:ls_encoding}
\end{figure*}%


\section{Evaluation} \label{sec:eval}

\begin{figure}[htb]
   \input{sections/longterm_prediction.tex}
\end{figure}%
\begin{figure}[htb]
   \input{sections/obstacle_generalization.tex}
   \vspace{-2.0em}
\end{figure}%
\begin{figure}[htb]
   \input{sections/inflow_generalization.tex}
\end{figure}%
In this section, we compare our architecture to the baseline of previous work.
We also perform an ablation study on different settings of our proposed architecture to compare their respective influence on the output.
We compute the mean peak signal-to-noise ratio (PSNR) for all our comparisons, i.e., larger values are better. For each case, we measure accuracy of our prediction w.r.t. density and velocity in terms of PSNR for ten simulations setups that were not seen during training.

For a thorough evaluation, we supply two prediction approaches.
First, we evaluate a regular prediction approach with no reinjection of physical information (denoted {\em VelDen}) that is in sync with previous work \cite{kim19a, lsp2019}.
This approach is formulated as
\begin{align} \label{eq:evalvelden}
   \tilde{\vc}^{t} = P(\tilde{\vc}^{t-2}, \tilde{\vc}^{t-1}),\\ \nonumber
   \tilde{\vx}^{t} = [\tilde{\vx}^{t}_{vel},\tilde{\vx}^{t}_{den}] = D(\tilde{\vc}^{t}),
\end{align}
where the previously predicted latent space points $\tilde{\vc}^{t-2}$ and $\tilde{\vc}^{t-1}$ are used to evaluate the next time step $\tilde{\vc}^{t}$.
Afterwards, $\tilde{\vc}^{t}$ is decoded $D(\tilde{\vc}^{t})$ and the density part $\tilde{\vx}^{t}_{den}$ is directly displayed, i.e., no external physical information about the system state is influencing the output of our {\em VelDen} benchmarks.

In the second approach, we make use of our LSS to reinject the advected density into the prediction to benefit from well understood physical computations that keep our predictions stable and can be performed in a fast manner.
The prediction process utilizing our LSS is denoted as \textit{Vel} and is given as
\begin{align} \label{eq:evalvel}
   \tilde{\vc}^{t} = P(\hat{\vc}^{t-2}, \hat{\vc}^{t-1}),\\ \nonumber
   \tilde{\vx}^{t} = [\tilde{\vx}^{t}_{vel},\tilde{\vx}^{t}_{den}] = D(\tilde{\vc}^{t}), \\ \nonumber
   \dot{\vx}^{t}_{den} = Adv(\dot{\vx}^{t-1}_{den}, \tilde{\vx}^{t}_{vel}),\\ \nonumber
   \dot{\vc}^{t} = E([\tilde{\vx}^{t}_{vel}, \dot{\vx}^{t}_{den}]),\\ \nonumber
   \hat{\vc}^{t} = [\tilde{\vc}^{t}_{vel}, \dot{\vc}^{t}_{den}],\\ \nonumber
   \hat{\vx}^{t} = [\tilde{\vx}^{t}_{vel},\dot{\vx}^{t}_{den}], \nonumber
\end{align}
where we are using the decoded predicted velocity $\tilde{\vx}^{t}_{vel}$ to advect the simulation density $\dot{\vx}^{t-1}_{den}$ and reinject its encoded form into our latent space $\hat{\vc}^{t}$.
The new latent space representation $\hat{\vc}^{t}$ is thereby formed by concatenating the new encoded density $ \dot{\vc}^{t}_{den}$ and the predicted encoded velocity field $\tilde{\vc}^{t}_{vel}$.
By reinjecting the advected density field $\dot{\vx}^{t}_{den}$, we inform the prediction network about boundary conditions as well as other known physical information that is external to the prediction state.
In the following we will ablate on different aspects of our method to evaluate their respective influence on the final results.

\textbf{Latent Space Temporal Awareness} \label{sec:tempaware_ls}
The temporal awareness of our spatial autoencoder is evaluated in this section, since it has a significant impact on
the performance of our temporal prediction network.
In \myreffig{ls_encoding} we evaluate three networks trained with different loss functions in terms of the stability of the latent space they generate for sequences.
For each of the plots,
$200$ frames of $20$ different smoke simulations were encoded to the latent space domain with an autoencoder.
The resulting latent space points were normalized with their respective maximum to the range of $[-1,1]$ and afterwards transformed to $3$ dimensions with PCA.
The supervised part was removed before normalization.
For our comparison we chose $3$ autoencoders with a latent space dimension of $16$.

The results in \myreffig{unordered_ls} were generated with a classic AE that was trained to only reproduce its input, i.e., only a direct loss on the output (\myrefeq{loss_AE}) was applied.
For this classic AE no temporal constraints were imposed, and no supervised parameters were added to the latent space.
The resulting PCA decomposition shows a very uneven distribution:
large distances between consecutive points exist in some siutations, whereas a large part of the samples are placed closely together in a cluster.

When adding
the supervised parameter loss (\myrefeq{loss_sup_params}) in the second evaluation~\cite{kim19a} the trajectories become more ordered, as shown in \myreffig{unordered_ls_sup_param}, but still exhibit a noticeably uneven distribution. Thus, the supervised parameter, despite being excluded from the PCA, has a noticeable influence on the latent space construction.

In \myreffig{ordered_ls}, the results of the AE trained with our proposed time-aware end-to-end training method are shown.
This AE applies the supervised parameter loss (\myrefeq{loss_sup_params}) as well as the direct loss on the output (\myrefeq{loss_AE}) and was trained in combination with the temporal prediction network P as described in \myrefsec{training}.
The visualization of the PCA decomposition shows that a strong temporal coherence of the consecutive data points emerges.
This visualization indicates why our prediction networks yield
substantial improvements in terms of accuracy: the end-to-end training 
provides the autoencoder with gradients that guide it towards learning 
a latent space mapping that is suitable for temporal predictions. Intuitively, changes over time require relatively small and smooth
updates, which results in the visually more regular curves shown in \myreffig{ordered_ls}.

\textbf{Simple LS Division vs. LSS} \label{sec:simplesplit_LSS}
A simple approach to arrive at a latent space with a clear subdivision in terms of input quantities is to use two separate spatial AEs for the individual input quantities.
After encoding, the two resulting latent space points $c_{vel} = E_{vel}(\vu)$ and $c_{den} = E_{den}(\rho)$ can be concatenated, yielding $c_{simple} = [c_{vel}, c_{den}]$.
In contrast to the simple approach, our LSS directly encodes both quantities with a single AE as $c_{LSS} = E([\vu,\rho])$ and enforces the subdivision with a soft-constraint.
The combined training with the prediction network is performed identical for both spatial compression versions.
It becomes apparent from the results in \myreftab{SS_LSS_vel_comparison} b), that the network trained with our soft-constraint outperforms the simple split variant.
Especially, when reinjecting the simulation density in the {\em Vel} benchmarks, we see a better PSNR value of $30.28$ for $\vu$ and $17.66$ for $\rho$ for our method in comparison to $26.95$ and $16.35$ for $\vu$ and $\rho$, respectively.
The reason for this is that the simple split version can not take advantage of synergistic effects in the input data, since both input quantities are encoded in separate AEs.
In contrast, our method uses the synergies of the physically interconnected velocity and density fields and robustly predicts future time steps.
\begin{figure}[htb]
   \begin{minipage}{\columnwidth}
      \begin{center}
         \footnotesize
         \captionsetup{type=table}
         \caption{ Evaluations for {\em Vel} and {\em VelDen} predictions; rotating and moving cup: a) Internal predictions $n_i$; b) Simple split vs. LSS; c) LS dimensionality comparison} \label{tab:IO_velden_comparison_rotcupmov} \label{tab:IO_vel_comparison_rotcupmov} \label{tab:SS_LSS_velden_comparison}  \label{tab:SS_LSS_vel_comparison} \label{tab:LSDim_vel_comparison} \label{tab:LSDim_velden_comparison}
         \resizebox{\tabularwidth\width}{!}{%
         \begin{tabular}{@{}>{\columncolor{white}[0pt][\tabcolsep]}cccc>{\columncolor{white}[\tabcolsep][0pt]}c@{}}
            \toprule
            \rowcolor{white}
            & \multicolumn{2}{c}{{\em Vel}} & \multicolumn{2}{c}{{\em VelDen}} \\
            \midrule
            \rowcolor{white}
            a) $n_i$ & PSNR $\vu$ & PSNR $\rho$ & PSNR $\vu$ & PSNR $\rho$ \\
            \midrule
            1 & 33.04 & 25.19 & 33.11 & 22.28 \\ \addlinespace
            6 & 35.89 & 26.28 & 36.07 & 25.41 \\ \addlinespace
            12 & \textbf{36.61} & \textbf{26.61} & \textbf{36.61} & \textbf{25.69} \\ 
            \midrule
            \midrule
            \rowcolor{white}
            \addlinespace
            b) Type & PSNR $\vu$ & PSNR $\rho$ & PSNR $\vu$ & PSNR $\rho$ \\
            \midrule
            Simple Split & 26.95 & 16.35 & \textbf{29.51} & 15.63 \\ \addlinespace
            LSS (ours) & \textbf{30.28} & \textbf{17.66} & 29.11 & \textbf{17.12}  \\ \addlinespace
            \midrule
            \midrule
            \rowcolor{white}
            \addlinespace
            c) LS Dimension $|c|$ & PSNR $\vu$ & PSNR $\rho$ & PSNR $\vu$ & PSNR $\rho$ \\ 
            \midrule
            16 & 30.28 & 17.66 &                        29.11 & 17.12 \\ \addlinespace
            32 & 30.40 & 17.64 &                        \textbf{31.58} & 18.90 \\ \addlinespace
            48 & \textbf{30.86} & \textbf{17.92} &      30.97 & \textbf{18.96} \\
            \bottomrule
         \end{tabular}
         }
         \global\rownum=0\relax
      \end{center}
   \end{minipage}%
   \vspace{-1.25em}
\end{figure}%

\textbf{Internal Iterations} \label{sec:internal_iterations}
We compare the internal iteration count of the prediction network in the training process in \myreftab{IO_vel_comparison_rotcupmov} a).
By performing multiple recurrent evaluations of our temporal network already in the training process we minimize the additive error build-up of many consecutive predictions.
To fight the additive prediction errors over a long time horizon is important to arrive at a robust and exact predicted sequence.
We chose the values of $1$, $6$ and $12$ internal iterations for our comparison.
It becomes apparent, that the network trained with $12$ internal iterations and thereby the longest prediction horizon is superior in both evaluations.
It should be noted, that the predictions with reinjection of the physical density field (\textit{Vel}) have a lower error on the density than the prediction-only (\textit{VelDen}) approach, e.g. a PSNR value of $26.61$ in contrast to $25.69$ for the density field of the $12$ iteration version.
This supports the usefulness of our proposed latent space subdivision, that is needed to reinject external information.

\textbf{Latent Space Dimensionality} \label{sec:lsdim}
The latent space dimensionality has a major impact on the resulting weight count of the autoencoder as well as the complexity of the temporal predictions and thereby their difficulty.
In the following we compare latent space sizes of $16$, $32$ and $48$.
When it comes to prediction only ({\em VelDen}), the PSNR is better for a larger latent space dimensionality.
In contrast to this observation, the PSNR value is on the same level for all latent space sizes, when the simulation density is reinjected ({\em Vel}).
For this reason we used a latent space dimensionality of $16$ for all further comparisons.
Due to bouyancy, the velocity and density of our smoke simulations are loosely coupled.
Thus, additional weights do not increase the overall performance when the reconstruction of the individual parts of the respective input quantities already converged.

\textbf{Latent Space Subdivision vs. No-Split} \label{sec:subdivision_nosplit}
To evaluate the usefulness of our LSS method that supports reinjection of external information, it is compared to a classic network setup that does not support that and instead performs a regular prediction.
When only performing a regular prediction, the step-wise prediction errors accumulate.
Without the reinjection of the externally driven density field into our prediction process, the quality of the outcome decreases drastically.
Due to the bouyancy based coupling of density and velocity this effect intensifies.
\begin{figure}[htb]
   \begin{minipage}{\columnwidth}
      \begin{center}
         \footnotesize
         \captionsetup{type=table}
         \caption{LSS and no-split comparison; rotating and moving cup; $400$ time steps}
         \label{tab:LSSplit_NoSplit_comparison_rotcupmov_longterm}
         \resizebox{\tabularwidth\width}{!}{%
         \begin{tabular}{@{}>{\columncolor{white}[0pt][\tabcolsep]}ccc>{\columncolor{white}[\tabcolsep][0pt]}c@{}}
            \toprule
            \rowcolor{white}
            LS Split & Type & PSNR $\vu$ & PSNR $\rho$ \\
            \midrule
            0.0 (no-split) & {\em VelDen} & 17.41 & 10.71 \\ \addlinespace
            0.66     & {\em VelDen} & 26.81 & 17.07 \\ \addlinespace
            0.66     & {\em Vel}    & \textbf{28.15} & \textbf{21.74} \\
            \bottomrule
         \end{tabular}
         }
         \global\rownum=0\relax
      \end{center}
   \end{minipage}%
   \vspace{-1.5em}
\end{figure}%

In \myreftab{LSSplit_NoSplit_comparison_rotcupmov_longterm} we compare the long-term temporal prediction performance of a $0.0$ (no-split) version and our $0.66$ LSS version over a time horizon of $400$ simulation steps.
Those split numbers correspond to the percentage designated for the velocity part of the latent space, e.g., a value of $0.66$ means that $66\%$ of the latent space are used for the encoded velocity.
Since the no-split version contains a classic AE without any latent space constraints, the  evaluation can only be performed for the prediction-only ({\em VelDen}) benchmark, since the reinjection of density is not possible with the classic AE.
Our LSS $0.66$ version with a density PSNR value of $21.74$ clearly outperforms the no-split version with a density PSNR value of $10.71$.
Due to the reinjection capabilities of our method, the resulting prediction remains stable whereas the classic no-split approach fails to capture the flow throughout the prediction horizon and even produces unphysical behavior (see \myreffig{longterm_prediction}).

\section{Results} \label{sec:results}
We demonstrate the effectiveness of the subdivided latent space with several generalization tests. 
As shown in \myreffig{obstacle_generalization}, our method is capable of predicting the fluid motion even when an obstacle is placed in the domain.
Due to our split latent space, the obstacle can be passively injected into the prediction process by supplying an encoded density field with a masked out obstacle region.
We replaced the density part of the latent space with its encoded state after advecting it with our predicted velocity field.
\begin{figure}[htb]%
   \input{sections/3d_results.tex}%
\end{figure}%
The prediction without injection of external information is not capturing the obstacle and thereby deviates from the ground truth.
In \myreffig{add_inflow_generalization} we show that our method is capable of predicting the fluid motion even when a new inflow region is added that was not seen while training.
Our network performs reasonably well in all tested experiments due to the reinjection capabilities provided by the latent space subdivision, even though the generalization scenes were not part of the training data.
\begin{center}
   \captionsetup{type=table}
   \caption{
   Average timing of a simulation step computed via regular pressure solve and our \textit{Vel} prediction scheme. 
   While the former scales with data complexity, ours scales linearly with the domain dimension. 
   Average of $5$ scenes with $100$ simulation steps each. Measured on Intel(R) Xeon(R) E5-1650 v3 (3.50GHz) and Nvidia GeForce RTX 2070.
   } \label{tab:perf_comp}
   \resizebox{\tabularwidth\width}{!}{%
   \begin{tabular}{@{}>{\columncolor{white}[0pt][\tabcolsep]}cccc>{\columncolor{white}[\tabcolsep][0pt]}c@{}}
      \toprule
      \rowcolor{white}
      Scene & Resolution & Type & Solve [s] & Total [s] \\
      \midrule
      Rot. mov. cup 3D & $48^3$ & Simulation & 0.891 & 0.960 \\ \addlinespace
      Rot. mov. cup 3D & $48^3$ & Prediction & \textbf{0.074} & \textbf{0.156} \\ \addlinespace
      \midrule
      Mov. smoke 3D & $48^3$     & Simulation &  0.472 & 0.537 \\ \addlinespace
      Mov. smoke 3D & $48^3$     & Prediction &  \textbf{0.059} & \textbf{0.132} \\ \addlinespace
      \midrule
      Rot. mov. cup & $64^2$     & Simulation & 0.041 & 0.044 \\ \addlinespace
      Rot. mov. cup & $64^2$     & Prediction & \textbf{0.012} & \textbf{0.019} \\ \addlinespace
      \midrule
      Rot. cup & $48^2$          & Simulation & 0.018 & 0.019 \\ \addlinespace
      Rot. cup & $48^2$          & Prediction & \textbf{0.011} & \textbf{0.015} \\
      \bottomrule
   \end{tabular}
   }
   \global\rownum=0\relax
\end{center}%
To demonstrate the capabilities of our method, we trained a 3D version of our network on the moving smoke scene; selected frames are shown in \myreffig{3d_result}. 
Additionally, we compared the runtime performance of our networks
to the regular solver that was used for generating the training data (see \myreftab{perf_comp}).
Even though, we need to decode and encode the density field due to our reinjection method and thereby copy it from GPU to CPU memory, we still arrive at a performance measure of $0.059$ seconds for an average prediction step in our 3D scene.
For comparison, a traditional multi-threaded CPU-based solver takes $0.472$ seconds on average for a simulation step for the same scenes.
%

\section{Conclusion \& Future Work} \label{sec:conclusion}
We have demonstrated an approach for subdividing latent spaces in a controlled manner, to improve generalization and long-term stability of physics predictions.
In combination with our time-aware end-to-end training that learns a reduced representation together with the time prediction, this makes it possible to predict sequences with several hundred roll-out steps.
In addition, our trained networks can be evaluated very efficiently, and yield significant speed-ups compared to traditional solvers.
As future work, we believe it will be interesting to further extend the generalizing capabilities of our network such that it can cover a wider range of physical behavior.
In addition, it will be interesting to explore different architectures to reduce the hardware requirements for training large 3D models with our approach.

\section*{Acknowledgements}
This work was supported by the ERC Starting Grant {\em realFlow} (StG\-2015\-637014) and  the Swiss National Science Foundation (grant no. 200021\_168997).
Source code and video: \url{https://ge.in.tum.de/publications/latent-space-subdivision/}.



\bibliography{paper}
\bibliographystyle{icml2020}


\newpage

\appendix
\newpage 
\ \\
\newpage

\begin{centering} \vspace{1cm}  \LARGE Supplemental Document for \ \\ Latent Space Subdivision: \ \\ Stable and Controllable Time Predictions for Fluid Flow  \\ \vspace{1cm}
\end{centering}

\normalsize

\section{Evaluation}
\begin{figure*}[ht]
    \newcommand\subfigwidth{.07\textwidth}
    \begin{center}
      \begin{subfigure}{\subfigwidth}
         \centering
         \begin{overpic}[width = 0.99\textwidth, trim=245 27 245 27, clip]{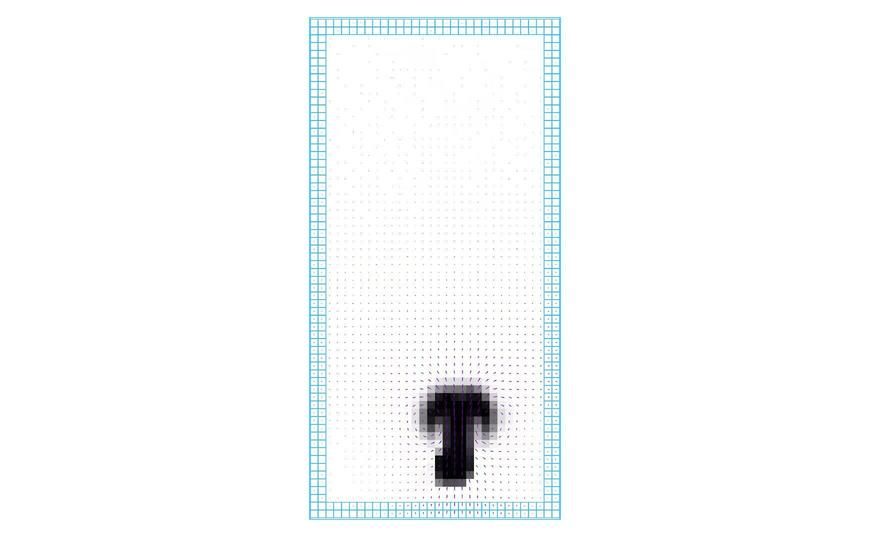}
            \put(0,80){\parbox{.5in}{\tiny Prediction only}}
         \end{overpic}
      \end{subfigure}%
      \begin{subfigure}{\subfigwidth}
         \centering
         \includegraphics[width = 0.99\textwidth, trim=245 27 245 27, clip]{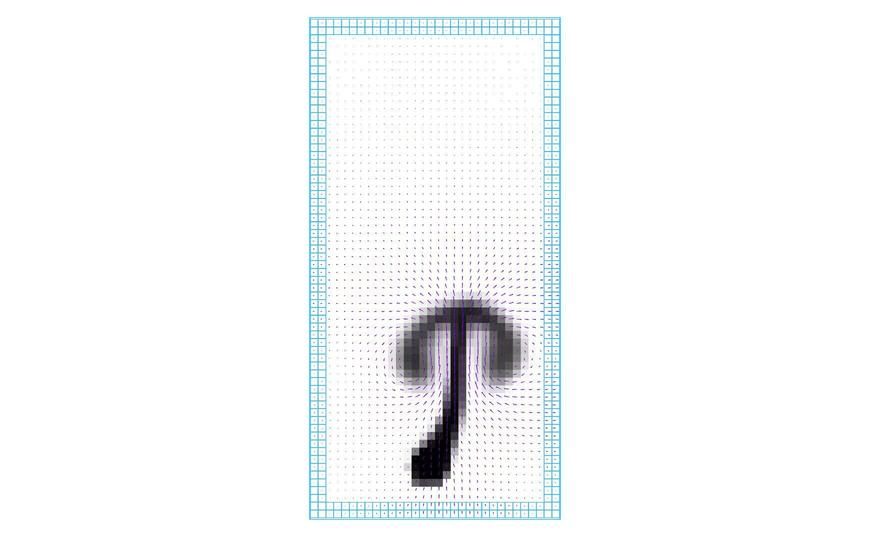}
       \end{subfigure}%
      \begin{subfigure}{\subfigwidth}
        \centering
        \includegraphics[width = 0.99\textwidth, trim=245 27 245 27, clip]{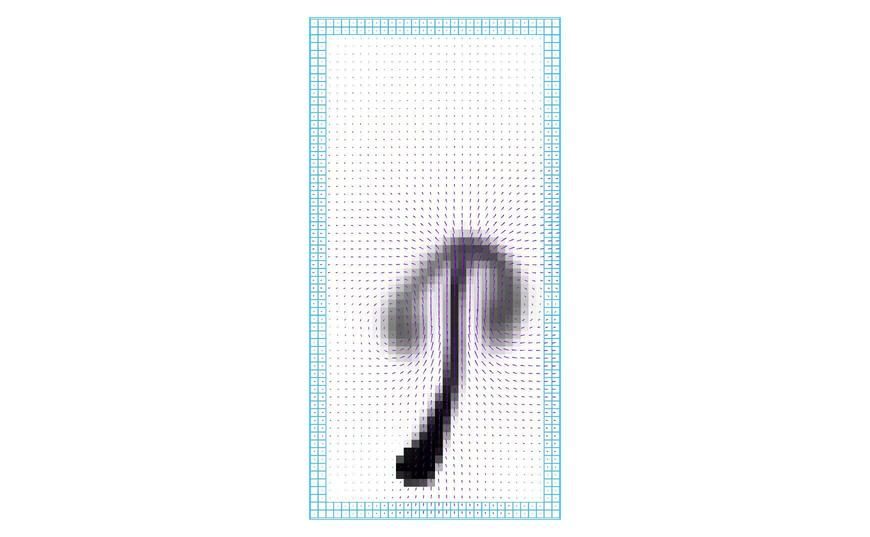}
      \end{subfigure}%
       \begin{subfigure}{\subfigwidth}
         \centering
         \includegraphics[width = 0.99\textwidth, trim=245 27 245 27, clip]{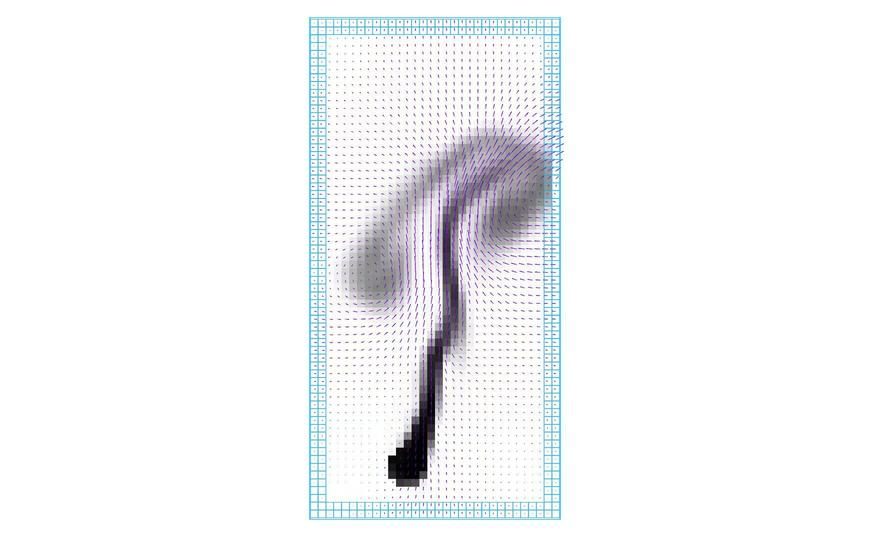}
       \end{subfigure}%
       \begin{subfigure}{\subfigwidth}
         \centering
         \includegraphics[width = 0.99\textwidth, trim=245 27 245 27, clip]{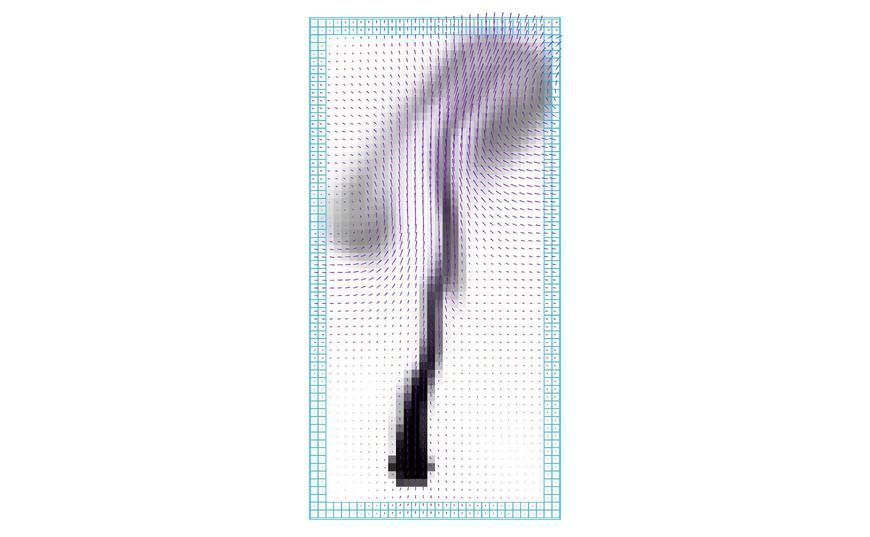}
       \end{subfigure}%
       \begin{subfigure}{\subfigwidth}
         \centering
         \includegraphics[width = 0.99\textwidth, trim=245 27 245 27, clip]{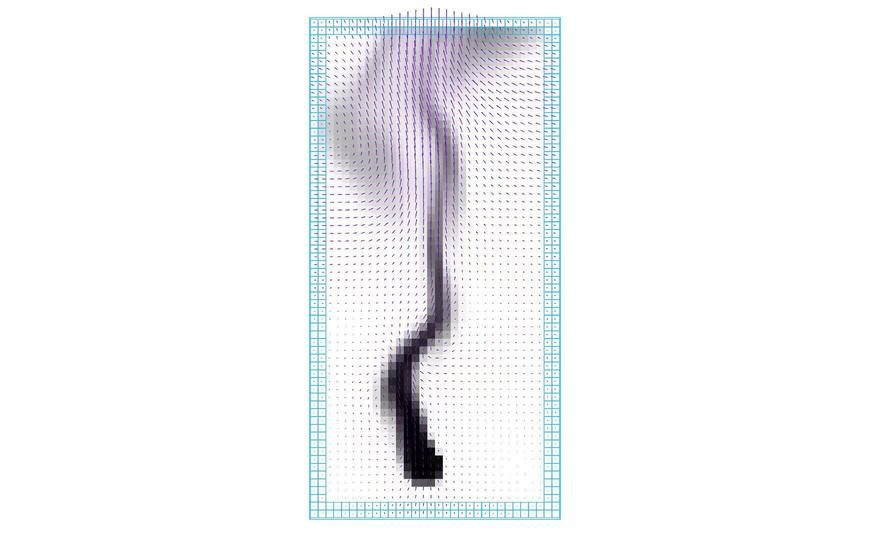}
       \end{subfigure}%
      \begin{subfigure}{\subfigwidth}
         \centering
         \includegraphics[width = 0.99\textwidth, trim=245 27 245 27, clip]{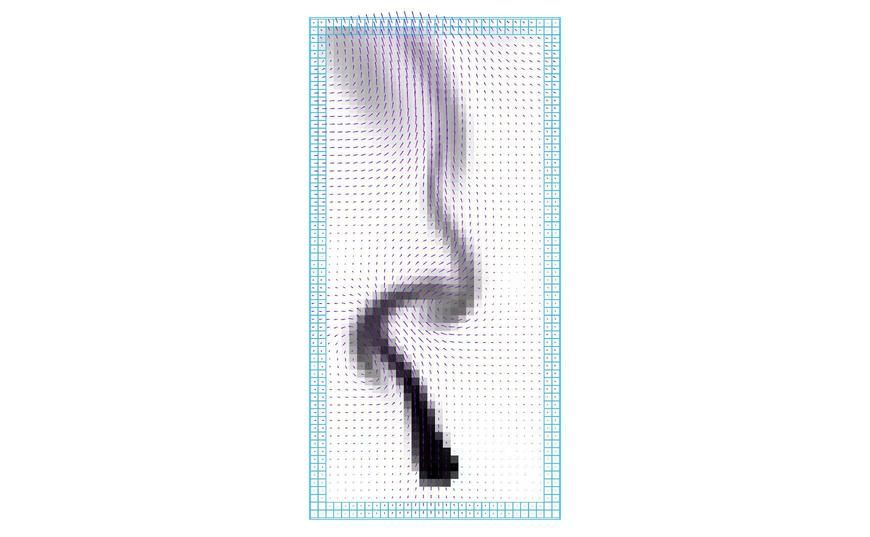}
      \end{subfigure}%
      \begin{subfigure}{\subfigwidth}
         \centering
         \includegraphics[width = 0.99\textwidth, trim=245 27 245 27, clip]{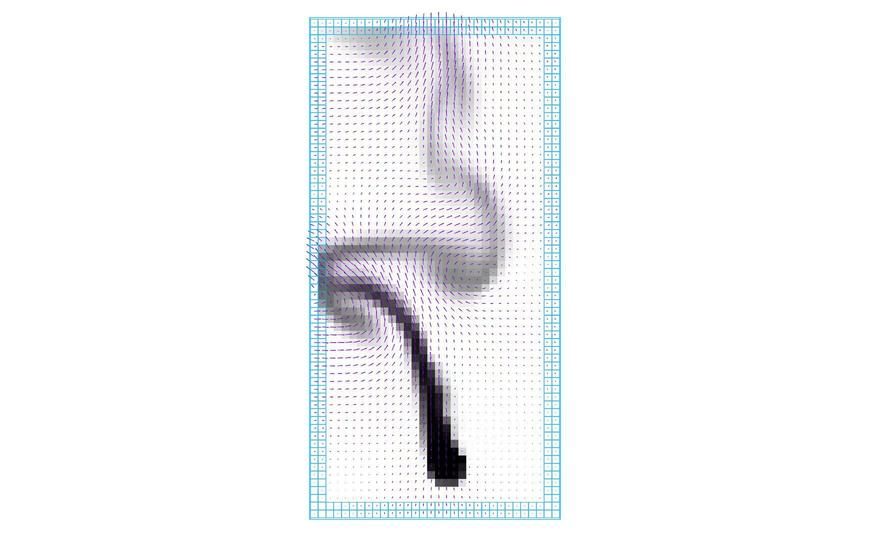}
      \end{subfigure}%
      \begin{subfigure}{\subfigwidth}
         \centering
         \includegraphics[width = 0.99\textwidth, trim=245 27 245 27, clip]{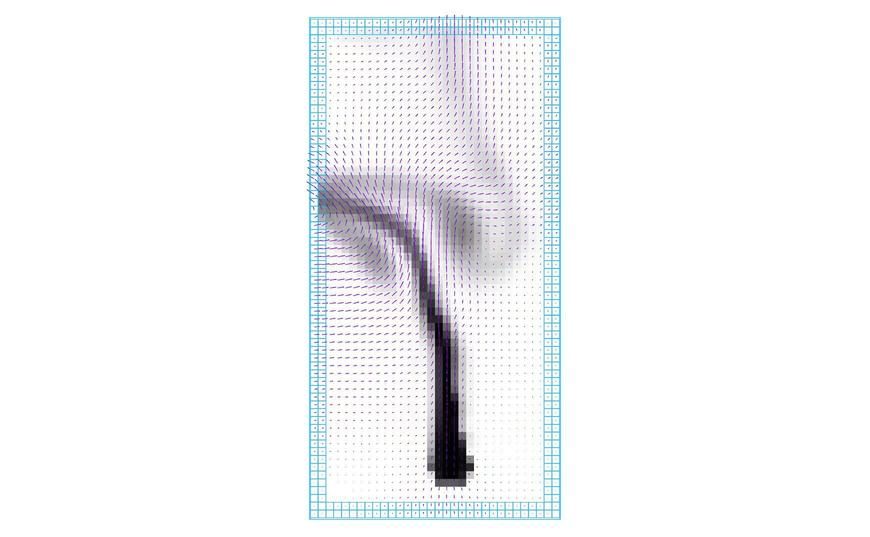}
      \end{subfigure}%
      \begin{subfigure}{\subfigwidth}
         \centering
         \includegraphics[width = 0.99\textwidth, trim=245 27 245 27, clip]{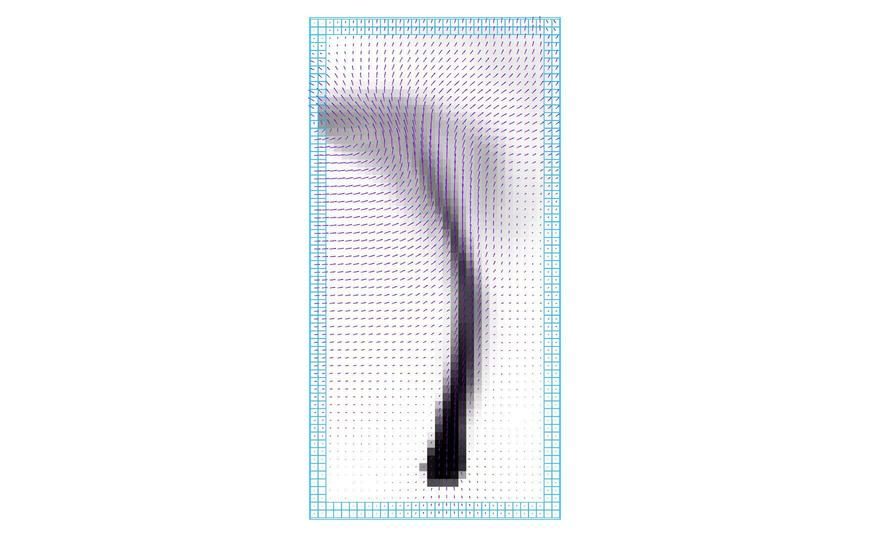}
      \end{subfigure}%
      \begin{subfigure}{\subfigwidth}
         \centering
         \includegraphics[width = 0.99\textwidth, trim=245 27 245 27, clip]{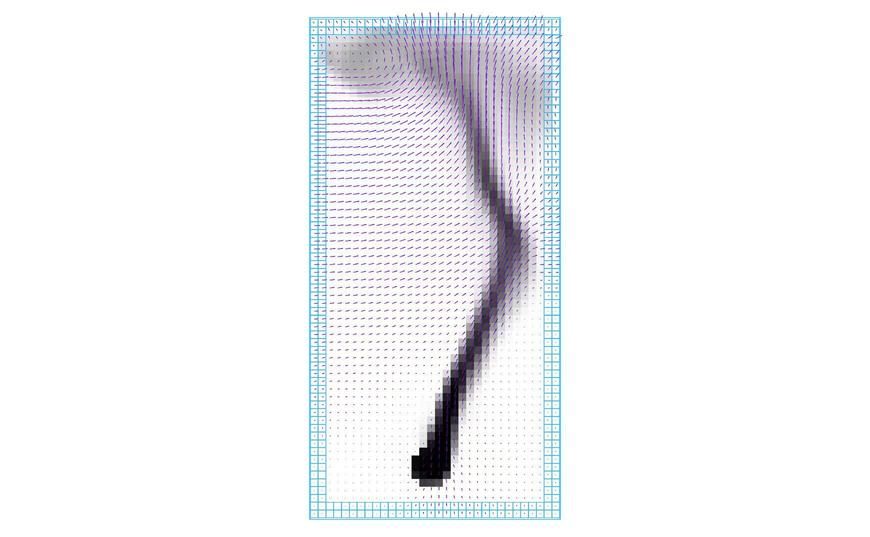}
      \end{subfigure}
   \end{center}
    \begin{center}
       \begin{subfigure}{\subfigwidth}
          \centering
          \begin{overpic}[width = 0.99\textwidth, trim=245 27 245 27, clip]{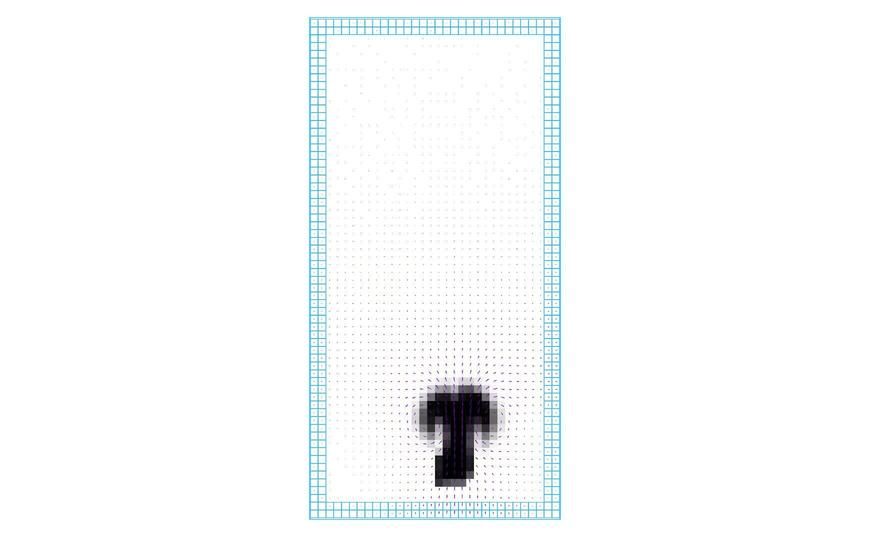}
             \put(0,80){\parbox{.5in}{\tiny Reinjection (ours)}}
          \end{overpic}
       \end{subfigure}%
       \begin{subfigure}{\subfigwidth}
          \centering
          \includegraphics[width = 0.99\textwidth, trim=245 27 245 27, clip]{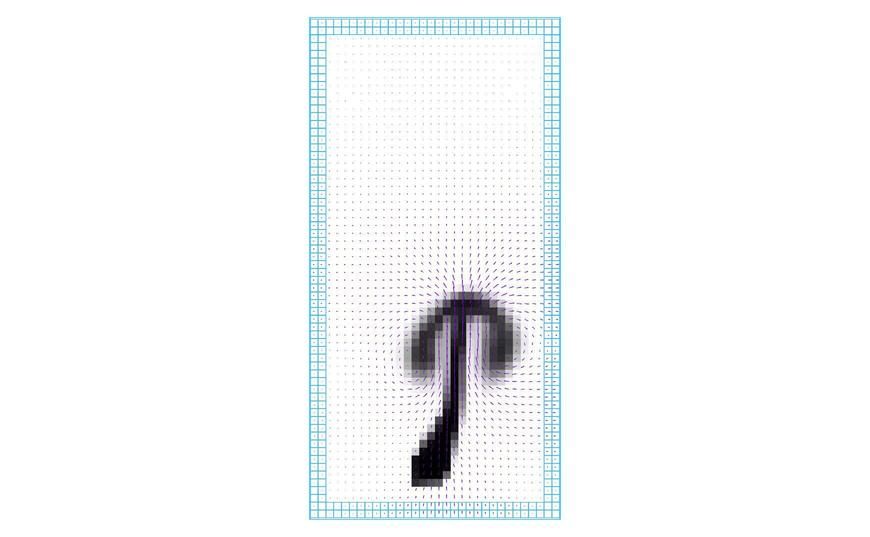}
       \end{subfigure}%
       \begin{subfigure}{\subfigwidth}
         \centering
         \includegraphics[width = 0.99\textwidth, trim=245 27 245 27, clip]{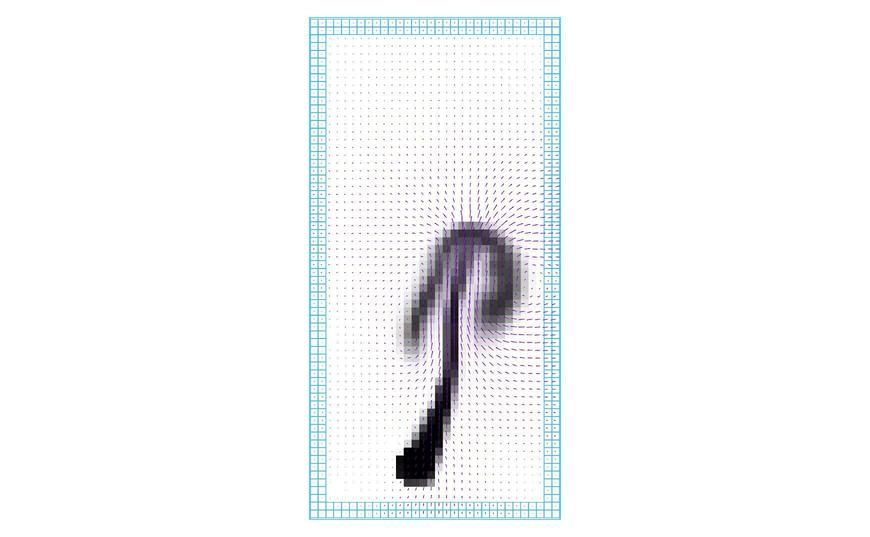}
       \end{subfigure}%
        \begin{subfigure}{\subfigwidth}
          \centering
          \includegraphics[width = 0.99\textwidth, trim=245 27 245 27, clip]{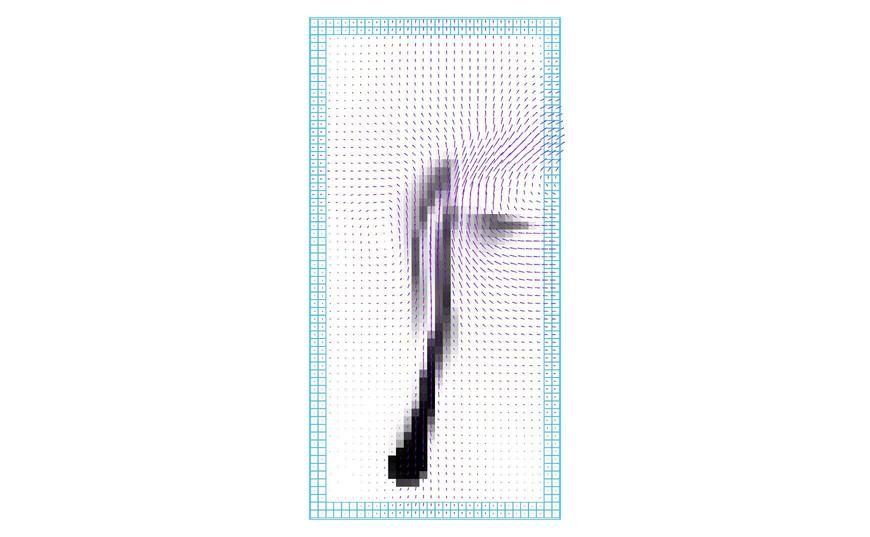}
        \end{subfigure}%
        \begin{subfigure}{\subfigwidth}
          \centering
          \includegraphics[width = 0.99\textwidth, trim=245 27 245 27, clip]{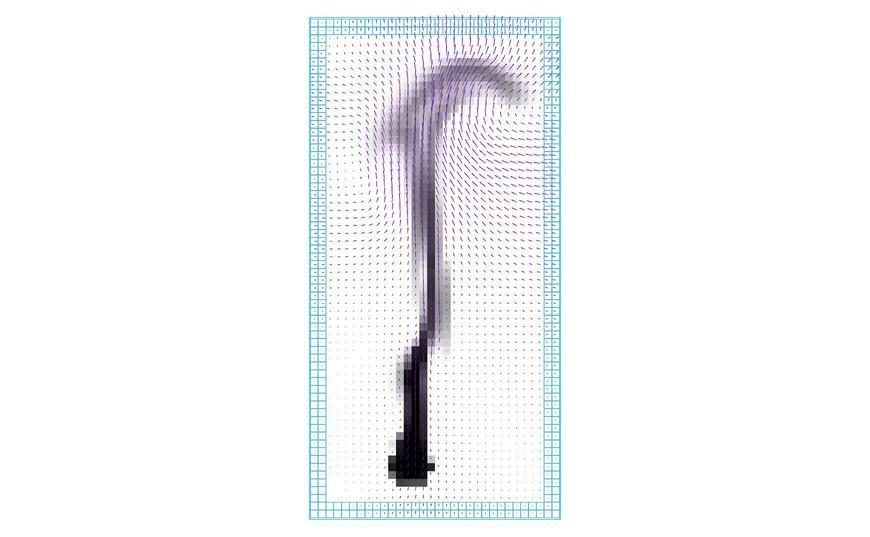}
        \end{subfigure}%
        \begin{subfigure}{\subfigwidth}
          \centering
          \includegraphics[width = 0.99\textwidth, trim=245 27 245 27, clip]{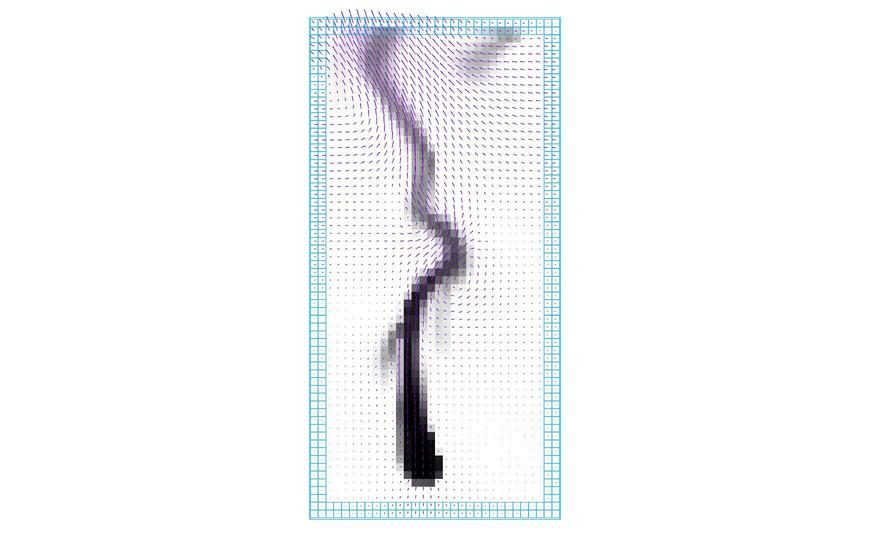}
        \end{subfigure}%
       \begin{subfigure}{\subfigwidth}
          \centering
          \includegraphics[width = 0.99\textwidth, trim=245 27 245 27, clip]{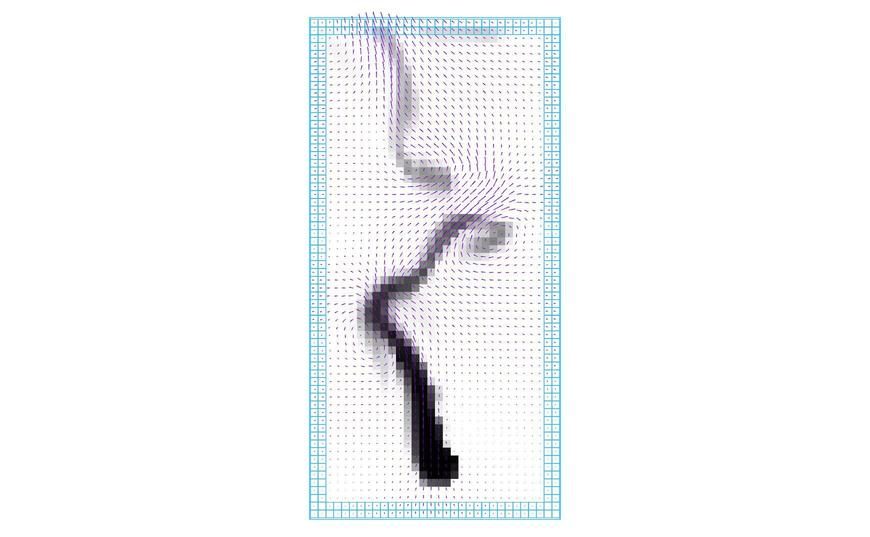}
       \end{subfigure}%
       \begin{subfigure}{\subfigwidth}
          \centering
          \includegraphics[width = 0.99\textwidth, trim=245 27 245 27, clip]{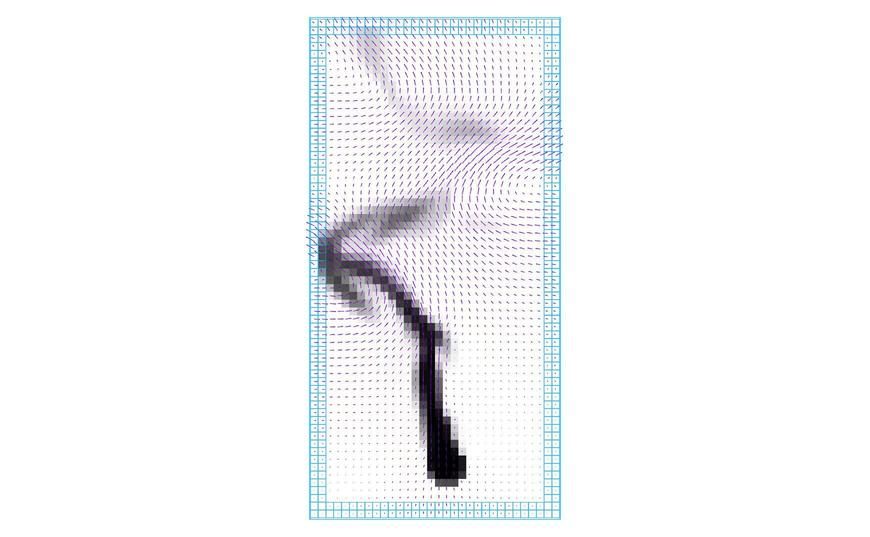}
       \end{subfigure}%
       \begin{subfigure}{\subfigwidth}
          \centering
          \includegraphics[width = 0.99\textwidth, trim=245 27 245 27, clip]{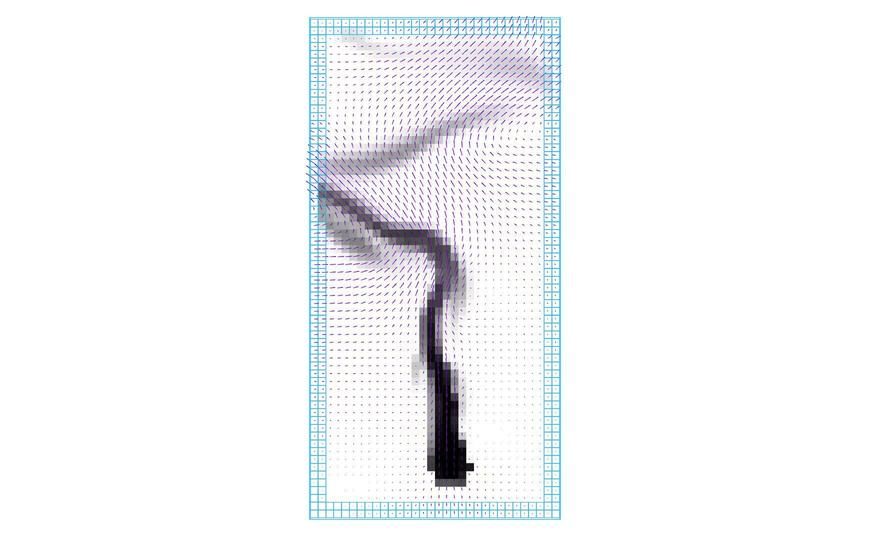}
       \end{subfigure}%
       \begin{subfigure}{\subfigwidth}
          \centering
          \includegraphics[width = 0.99\textwidth, trim=245 27 245 27, clip]{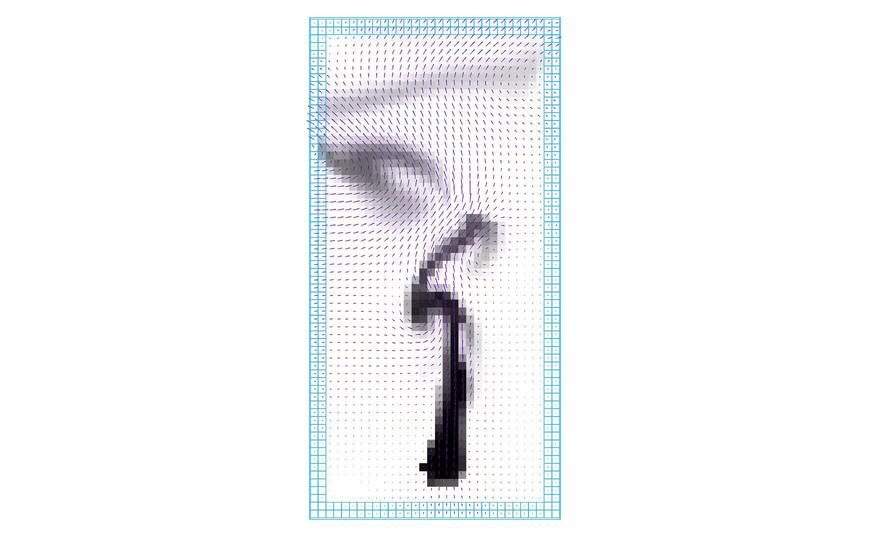}
       \end{subfigure}%
       \begin{subfigure}{\subfigwidth}
          \centering
          \includegraphics[width = 0.99\textwidth, trim=245 27 245 27, clip]{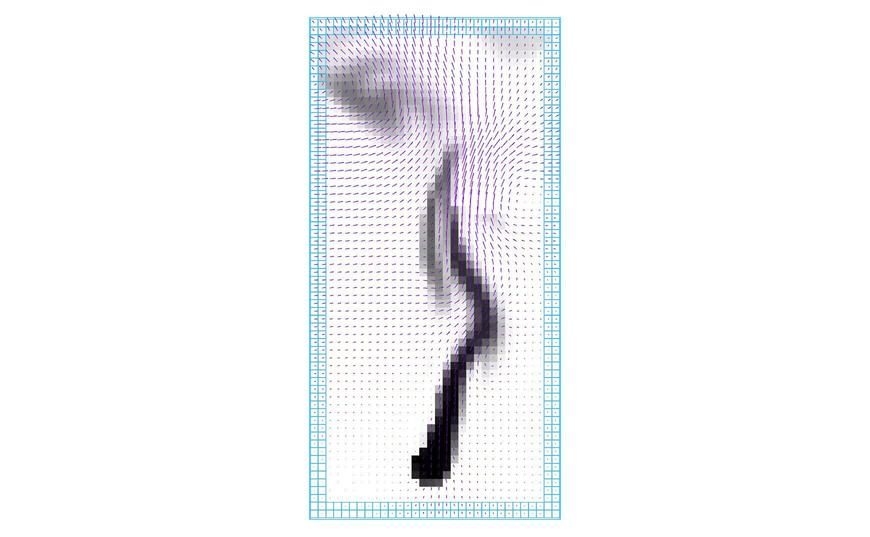}
       \end{subfigure}
    \end{center}
    \begin{center}
       \begin{subfigure}{\subfigwidth}
          \centering
          \begin{overpic}[width = 0.99\textwidth, trim=245 27 245 27, clip]{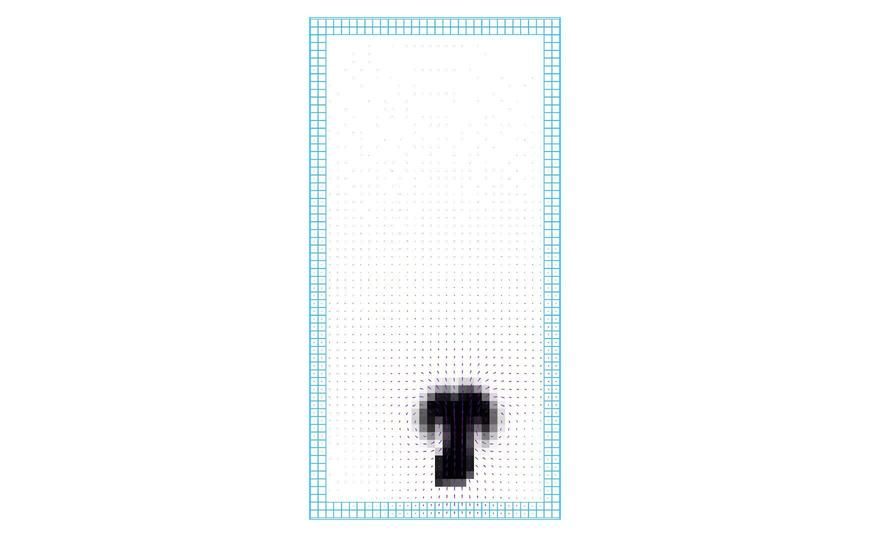}
             \put(0,80){\tiny GT}
           \end{overpic}
       \end{subfigure}%
       \begin{subfigure}{\subfigwidth}
          \centering
          \includegraphics[width = 0.99\textwidth, trim=245 27 245 27, clip]{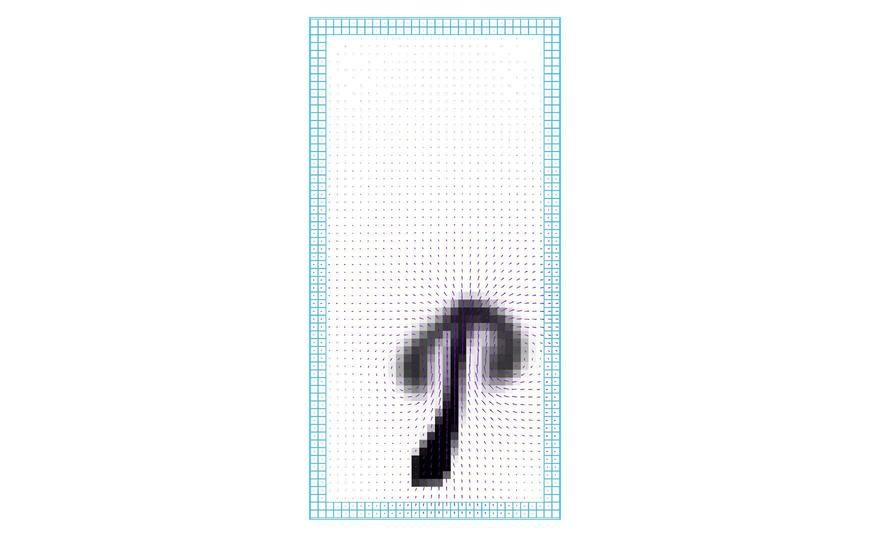}
        \end{subfigure}%
       \begin{subfigure}{\subfigwidth}
         \centering
         \includegraphics[width = 0.99\textwidth, trim=245 27 245 27, clip]{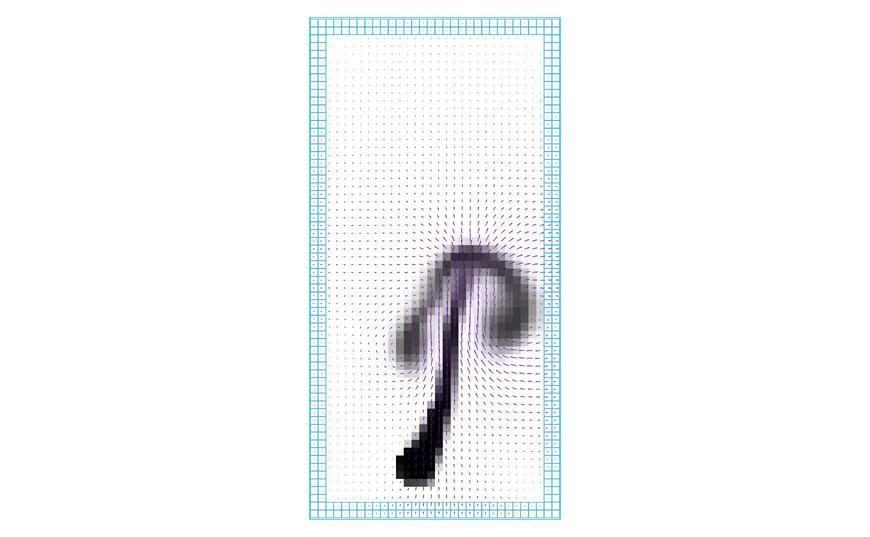}
       \end{subfigure}%
        \begin{subfigure}{\subfigwidth}
          \centering
          \includegraphics[width = 0.99\textwidth, trim=245 27 245 27, clip]{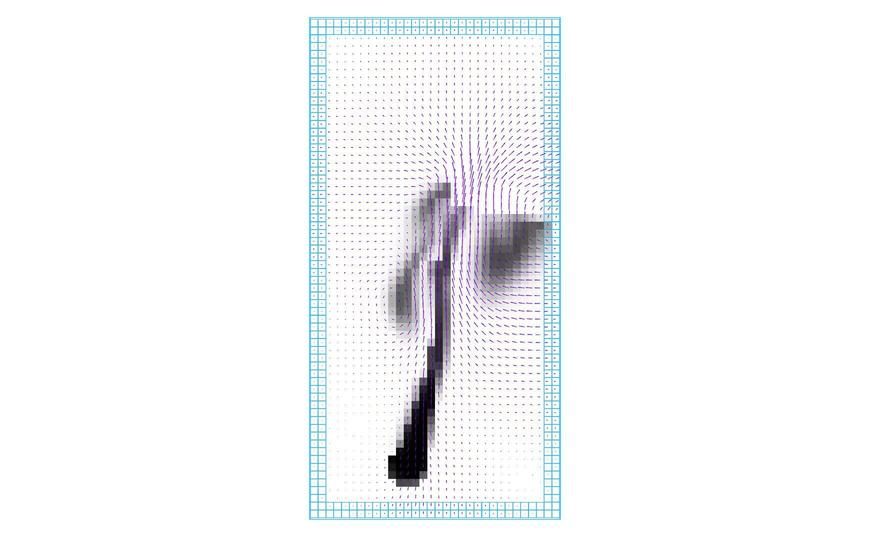}
        \end{subfigure}%
        \begin{subfigure}{\subfigwidth}
          \centering
          \includegraphics[width = 0.99\textwidth, trim=245 27 245 27, clip]{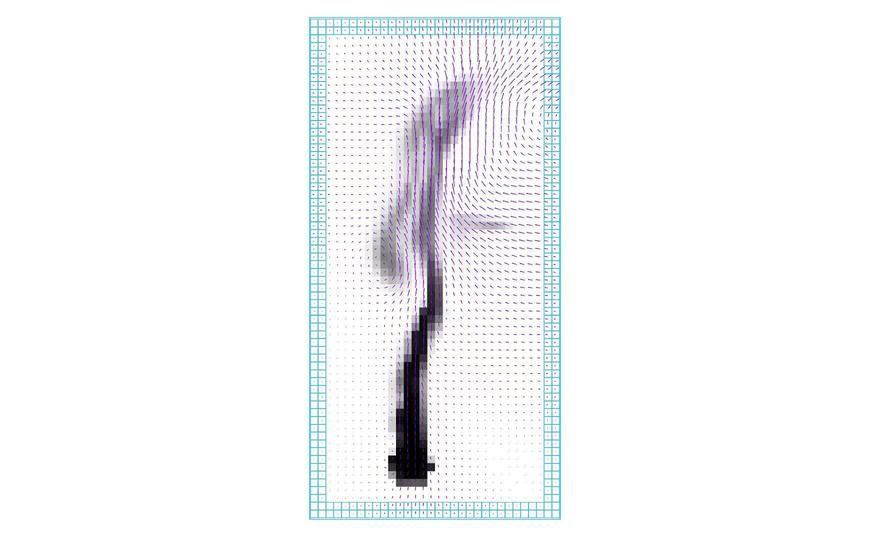}
        \end{subfigure}%
        \begin{subfigure}{\subfigwidth}
          \centering
          \includegraphics[width = 0.99\textwidth, trim=245 27 245 27, clip]{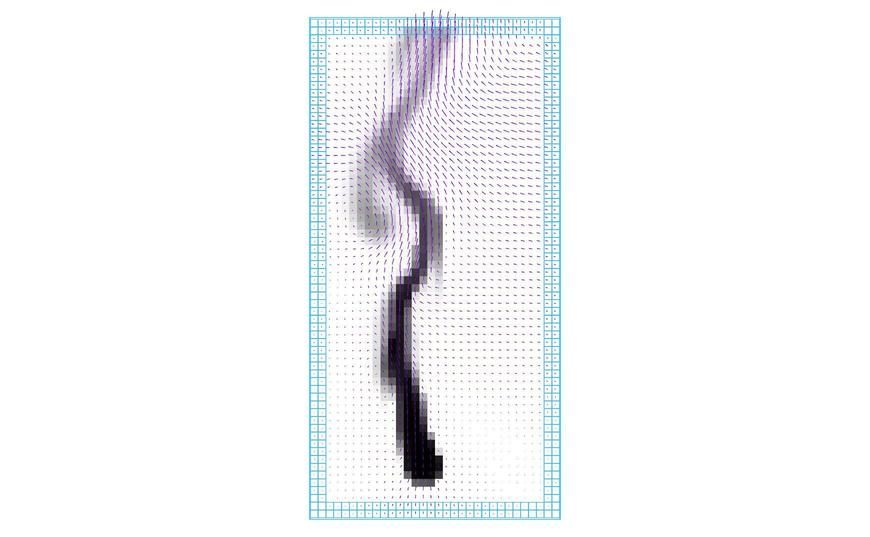}
        \end{subfigure}%
       \begin{subfigure}{\subfigwidth}
          \centering
          \includegraphics[width = 0.99\textwidth, trim=245 27 245 27, clip]{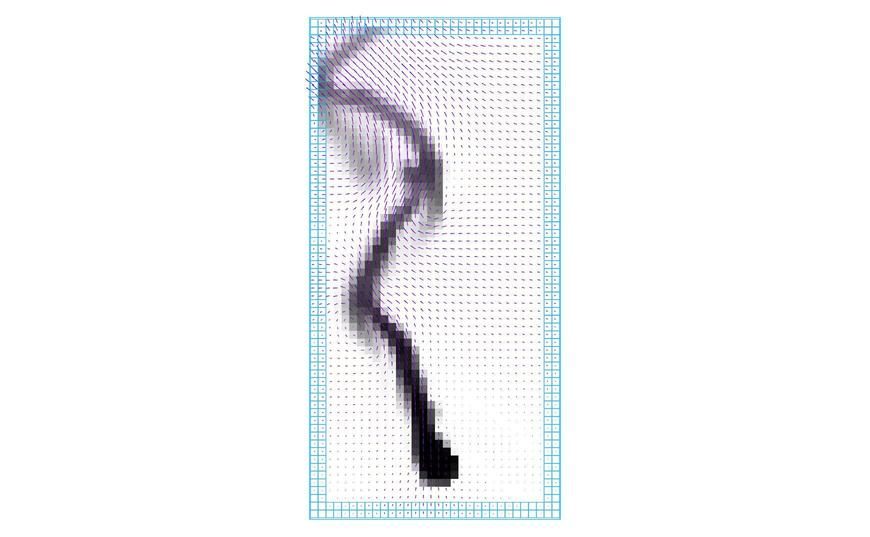}
       \end{subfigure}%
       \begin{subfigure}{\subfigwidth}
          \centering
          \includegraphics[width = 0.99\textwidth, trim=245 27 245 27, clip]{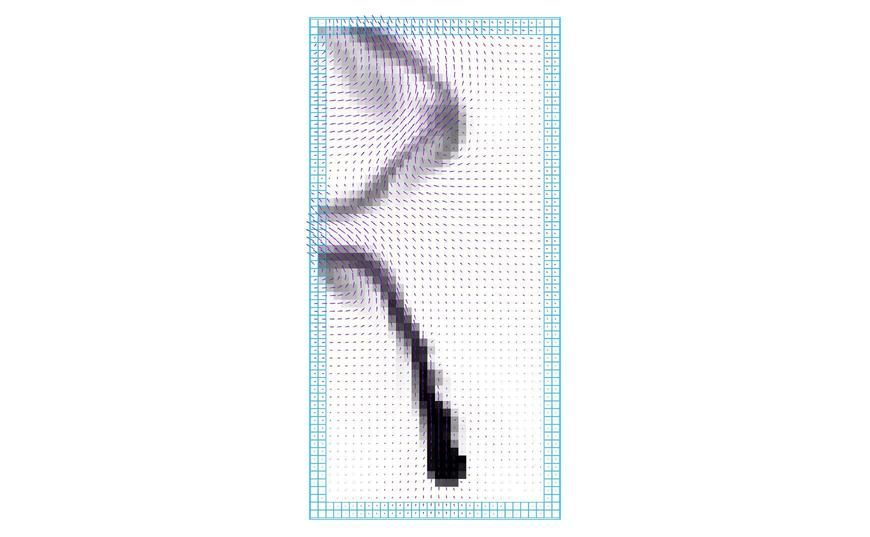}
       \end{subfigure}%
       \begin{subfigure}{\subfigwidth}
          \centering
          \includegraphics[width = 0.99\textwidth, trim=245 27 245 27, clip]{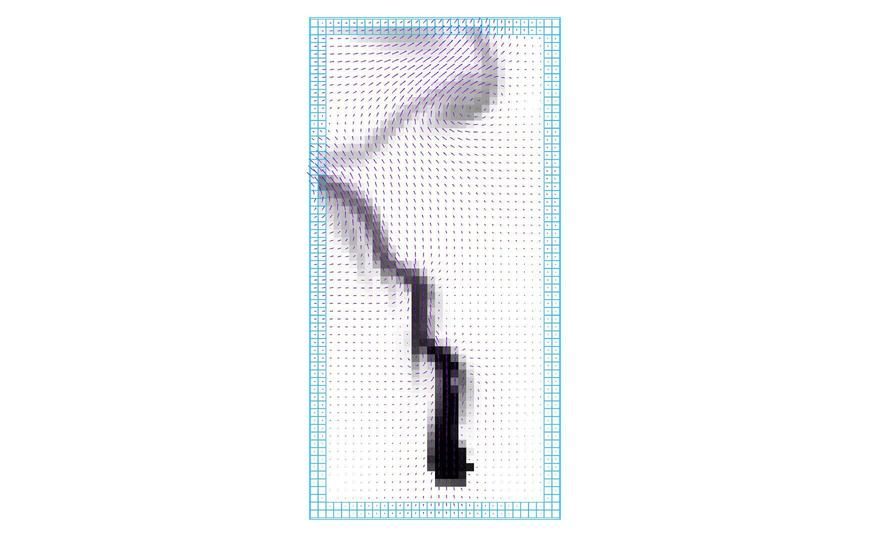}
       \end{subfigure}%
       \begin{subfigure}{\subfigwidth}
          \centering
          \includegraphics[width = 0.99\textwidth, trim=245 27 245 27, clip]{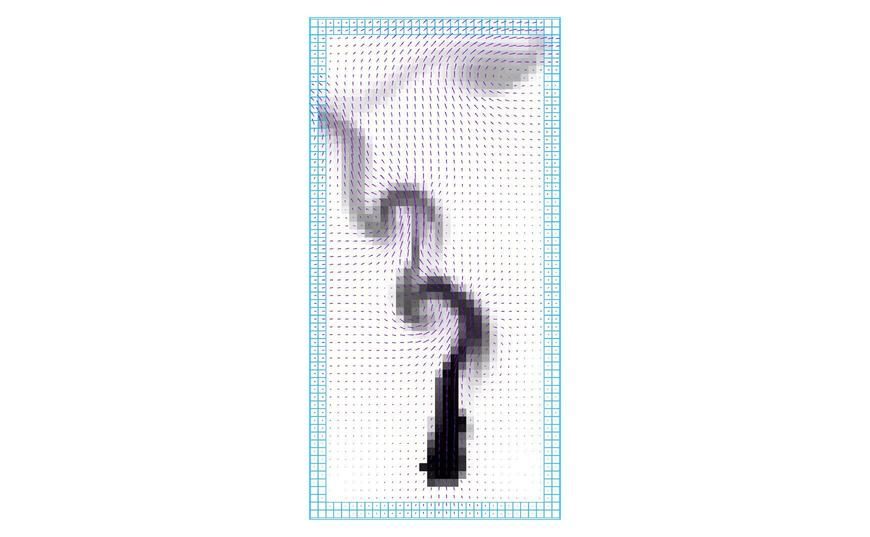}
       \end{subfigure}%
       \begin{subfigure}{\subfigwidth}
          \centering
          \includegraphics[width = 0.99\textwidth, trim=245 27 245 27, clip]{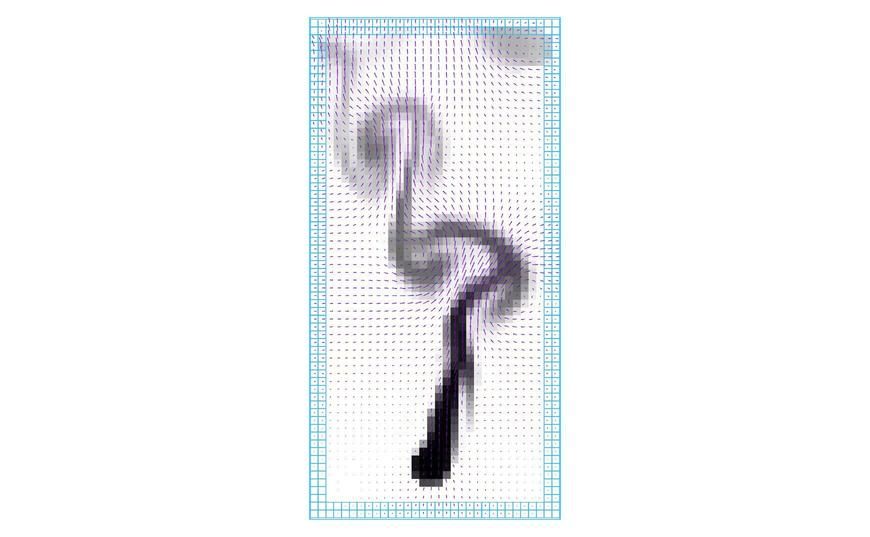}
       \end{subfigure}
    \end{center}
    \caption{ An sink is placed in the upper right of our moving smoke scene. This was unseen during training. The prediction by our proposed method remains stable and realistic. In the second row density reinjection was applied. In the top row no external information was injected. Thus, the sink can't be processed by the network. }
    \label{fig:sink_generalization}
 \end{figure*}

\subsection{Prediction Window Size} \label{sec:ap_predwindow}

The prediction window $w$ describes the count of consecutive time steps that are taken as input by the temporal prediction network.
In our comparison we tested window sizes ranging from $2$ over $3$ up to $4$ consecutive input steps.
The results in terms of PSNR values are displayed in \myreftab{PredWindow_vel_comparison} and \myreftab{PredWindow_velden_comparison}.
\begin{minipage}{0.24\textwidth}
   \begin{center}
      \captionsetup{type=table}
      \caption{Prediction window $w$ comparison {\em Vel}} \label{tab:PredWindow_vel_comparison}
      \resizebox{\tabularwidth\width}{!}{%
      \begin{tabular}{@{}>{\columncolor{white}[0pt][\tabcolsep]}cc>{\columncolor{white}[\tabcolsep][0pt]}c@{}}
         \toprule
         \rowcolor{white}
         $w$ & PSNR $\vu$ & PSNR $\rho$ \\
         \midrule
         4 & 29.67 & 17.04 \\ \addlinespace
         3 & 29.79 & 16.87 \\ \addlinespace
         2 & \textbf{30.28} & \textbf{17.66} \\
         \bottomrule
      \end{tabular}
      }
      \global\rownum=0\relax
   \end{center}
\end{minipage}%
\begin{minipage}{0.24\textwidth}
   \begin{center}
      \captionsetup{type=table}
      \caption{Prediction window $w$ comparison {\em VelDen}} \label{tab:PredWindow_velden_comparison}
      \resizebox{\tabularwidth\width}{!}{%
      \begin{tabular}{@{}>{\columncolor{white}[0pt][\tabcolsep]}cc>{\columncolor{white}[\tabcolsep][0pt]}c@{}}
         \toprule
         \rowcolor{white}
         $w$ & PSNR $\vu$ & PSNR $\rho$ \\
         \midrule
         4 & \textbf{29.98} & \textbf{18.24} \\ \addlinespace
         3 & 29.52 & 17.84 \\ \addlinespace
         2 & 29.11 & 17.12 \\
         \bottomrule
      \end{tabular}
      }
      \global\rownum=0\relax
   \end{center}
\end{minipage}%

It becomes apparent that the prediction-only approach (\textit{VelDen}) benefits from a larger input window, whereas the \textit{Vel} approach with reinjected external information performs best with a smaller input window.

\subsection{Latent Space Split Percentage} \label{sec:ap_lssplit}

We evaluated the impact of the latent space split percentage on three of our datasets.
Therefore, we trained multiple models with different split percentages on the individual datasets.
The comparison for our moving smoke scene is shown in \myreftab{LSSplit_vel_comparison_movsmoke} and \myreftab{LSSplit_velden_comparison_movsmoke}.
The latter are the results of the prediction-only evaluation (denoted {\em VelDen}), whereas the first table presents the results of our reinjected density approach (denoted {\em Vel}).
In this experiment all split versions are outperformed by the no-split version in the prediction-only only setup with PSNR values of $29.71$ and $18.03$ for velocity and density, respectively.

\begin{center}
    \begin{minipage}{0.24\textwidth}
       \begin{center}
          \captionsetup{type=table}
          \caption{LS split comparison {\em Vel}; moving smoke} \label{tab:LSSplit_vel_comparison_movsmoke}
          \resizebox{\tabularwidth\width}{!}{%
          \begin{tabular}{@{}>{\columncolor{white}[0pt][\tabcolsep]}cc>{\columncolor{white}[\tabcolsep][0pt]}c@{}}
             \toprule
             \rowcolor{white}
             LS Split & PSNR $\vu$ & PSNR $\rho$ \\
             \midrule
             0.33 & 28.07 & 15.84 \\ \addlinespace
             0.5  & 29.28 & 16.49 \\ \addlinespace
             0.66 & \textbf{30.28} & \textbf{17.66} \\
             \bottomrule
          \end{tabular}
          }
          \global\rownum=0\relax
       \end{center}
    \end{minipage}%
    \begin{minipage}{0.24\textwidth}
       \begin{center}
          \captionsetup{type=table}
          \caption{LS split comparison {\em VelDen}; moving smoke} \label{tab:LSSplit_velden_comparison_movsmoke}
          \resizebox{\tabularwidth\width}{!}{%
          \begin{tabular}{@{}>{\columncolor{white}[0pt][\tabcolsep]}cc>{\columncolor{white}[\tabcolsep][0pt]}c@{}}
             \toprule
             \rowcolor{white}
             LS Split & PSNR $\vu$ & PSNR $\rho$ \\
             \midrule
             0.0 (no-split) & \textbf{29.71} & \textbf{18.03} \\ \addlinespace
             0.33 & 28.49 & 16.63 \\ \addlinespace
             0.5  & 29.06 & 17.38 \\ \addlinespace
             0.66 & 29.11 & 17.12 \\ \addlinespace
             \bottomrule
          \end{tabular}
          }
          \global\rownum=0\relax
       \end{center}
    \end{minipage}%
\end{center}

\begin{center}
    \begin{minipage}{0.24\textwidth}
       \begin{center}
          \captionsetup{type=table}
          \caption{LS split comparison {\em Vel}; rotating cup} \label{tab:LSSplit_vel_comparison_rotcup}
          \resizebox{\tabularwidth\width}{!}{%
          \begin{tabular}{@{}>{\columncolor{white}[0pt][\tabcolsep]}cc>{\columncolor{white}[\tabcolsep][0pt]}c@{}}
             \toprule
             \rowcolor{white}
             LS Split & PSNR $\vu$ & PSNR $\rho$ \\
             \midrule
             0.33 & 36.67 & 28.46 \\ \addlinespace
             0.5  & 36.66 & 29.22 \\ \addlinespace
             0.66 & \textbf{38.52} & \textbf{29.73} \\
             \bottomrule
          \end{tabular}
          }
          \global\rownum=0\relax
       \end{center}
    \end{minipage}%
    \begin{minipage}{0.24\textwidth}
       \begin{center}
          \captionsetup{type=table}
          \caption{LS split comparison {\em VelDen}; rotating cup} \label{tab:LSSplit_velden_comparison_rotcup}
          \resizebox{\tabularwidth\width}{!}{%
          \begin{tabular}{@{}>{\columncolor{white}[0pt][\tabcolsep]}cc>{\columncolor{white}[\tabcolsep][0pt]}c@{}}
             \toprule
             \rowcolor{white}
             LS Split & PSNR $\vu$ & PSNR $\rho$ \\
             \midrule
             0.0 (no-split) & \textbf{37.90} & 22.68 \\ \addlinespace
             0.33 & 37.52 & 25.01 \\ \addlinespace
             0.5  & 36.77 & 25.13 \\ \addlinespace
             0.66 & 37.57 &  \textbf{25.32} \\
             \bottomrule
          \end{tabular}
          }
          \global\rownum=0\relax
       \end{center}
    \end{minipage}%
\end{center}

\begin{center}
    \begin{minipage}{0.24\textwidth}
       \begin{center}
          \captionsetup{type=table}
          \caption{LS split comparison {\em Vel}; rotating and moving cup} \label{tab:LSSplit_vel_comparison_rotcupmov}
          \resizebox{\tabularwidth\width}{!}{%
          \begin{tabular}{@{}>{\columncolor{white}[0pt][\tabcolsep]}cc>{\columncolor{white}[\tabcolsep][0pt]}c@{}}
             \toprule
             \rowcolor{white}
             LS Split & PSNR $\vu$ & PSNR $\rho$ \\
             \midrule
             0.33 & 35.67 & 25.10 \\ \addlinespace
             0.5  & \textbf{36.94} & \textbf{26.59} \\ \addlinespace
             0.66 & 36.50 & 26.25 \\
             \bottomrule
          \end{tabular}
          }
          \global\rownum=0\relax
       \end{center}
    \end{minipage}%
    \begin{minipage}{0.24\textwidth}
       \begin{center}
          \captionsetup{type=table}
          \caption{LS split comparison {\em VelDen}; rotating and moving cup} \label{tab:LSSplit_velden_comparison_rotcupmov}
          \resizebox{\tabularwidth\width}{!}{%
          \begin{tabular}{@{}>{\columncolor{white}[0pt][\tabcolsep]}cc>{\columncolor{white}[\tabcolsep][0pt]}c@{}}
             \toprule
             \rowcolor{white}
             LS Split & PSNR $\vu$ & PSNR $\rho$ \\
             \midrule
             0.33 & \textbf{37.89} & \textbf{26.45} \\ \addlinespace
             0.5  & 37.30 & 26.16 \\ \addlinespace
             0.66 & 37.56 &  26.14 \\ 
             \bottomrule
          \end{tabular}
          }
          \global\rownum=0\relax
       \end{center}
    \end{minipage}%
  
    \begin{minipage}{0.24\textwidth}
      \begin{center}
         \footnotesize
         \captionsetup{type=table}
         \caption{LSS and no-split comparison; rotating cup; $100$ time steps} \label{tab:LSSplit_NoSplit_comparison_rotcup}
         \resizebox{\tabularwidth\width}{!}{%
         \begin{tabular}{@{}>{\columncolor{white}[0pt][\tabcolsep]}ccc>{\columncolor{white}[\tabcolsep][0pt]}c@{}}
            \toprule
            \rowcolor{white}
            LS Split & Type & PSNR $\vu$ & PSNR $\rho$ \\
            \midrule
            0.0 (no-split) & {\em VelDen} & 37.90 & 22.68 \\ \addlinespace
            0.66     & {\em VelDen} & 37.57 & 25.32 \\ \addlinespace
            0.66     & {\em Vel}    & \textbf{38.52} & \textbf{29.73} \\
            \bottomrule
         \end{tabular}
         }
         \global\rownum=0\relax
      \end{center}
   \end{minipage}%
\end{center}

In contrast, the networks trained on the rotating cup dataset behave different as shown in \myreftab{LSSplit_vel_comparison_rotcup} and \myreftab{LSSplit_velden_comparison_rotcup}.
The classic no-split version is outperformed by all other split versions in terms of density PSNR values in the prediction-only ({\em VelDen}) setup.
In the reinjected density evaluation ({\em Vel}), the benefit of latent space splitting becomes even more apparent when comparing the PSNR values of velocity, $38.52$ and density, $29.73$ of the $0.66$ network with the velocity PSNR of $37.90$ and density PSNR $22.68$ of the no-split version.

\subsection{Latent Space Subdivision vs. No-Split} \label{sec:ap_subdivision_nosplit}
In this section we present additional results for our rotating cup dataset. See the main document for a long-term comparison of LSS vs. no-split for our more complicated moving and rotating cup dataset.
In \myreftab{LSSplit_NoSplit_comparison_rotcup} we compare the temporal prediction performance of a $0.0$ (no-split) version and our $0.66$ LSS version over a time horizon of $100$ simulation steps.
Our LSS $0.66$ version with a density PSNR value of $29.73$ clearly outperforms the no-split version with a density PSNR value of $22.68$.

\subsection{Generalization} \label{sec:ap_generalization}
Additionally, we show in \myreffig{sink_generalization} that our method recovers from the removal of smoke in a certain sink-region and is capable of predicting the fluid motion. 

\section{Fluid Flow Data}%
Our work concentrates on single-phase flows, modelled by a pressure-velocity formulation of the incompressible Navier-Stokes equations as highlighted in \myrefsec{relwork}.
Thereby, we apply a classic NS solver to simulate our smoke flows based on R. Bridson \yrcite{bridson2015}.
In addition to \myrefsec{training}, more information about the simulation procedure is provided in the following.%

\subsection{Simulation Setup}
The linear system for pressure projection is solved with 
a conjugate gradient method.
The conjugate gradient (CG) solver accuracy is set to $1\cdot10^{-4}$ for our moving smoke dataset, whereas an accuracy of $1\cdot10^{-3}$ is utilized for the moving cup datasets.
We generated all our datasets with a time step of $0.5$. 
Depending on the behavioral requirements of our different experiments with rising, hot and sinking, cold smoke we 
use the Boussinesq model with the smoke density
in combination with a gravity constant of $(0.0, -4\cdot10^{-3}, 0.0)$ for the moving and rising smoke and $(0.0, 1\cdot10^{-3}, 0.0)$ for the rotating cup dataset.
To arrive at a more turbulent flow behavior, the gravity constant was set to $(0.0, 1\cdot10^{-2}, 0.0)$ for our moving and rotating cup dataset.
We do not apply other forces or additional viscosity. We purely rely on numerical diffusion to introduce viscosity effects.

In combination with the quantities required by our classic NS setup, namely flow velocity $\vu$, pressure $p$ and density $\rho$, we also need a flag grid $f$, an obstacle velocity field $\vu_{obs}$ and the corresponding obstacle levelset for our obstacle supporting scenes.
Thereby our density $\rho$ is passively advected within the flow velocity $\vu$.

To handle the obstacle movement accordingly, we calculate the obstacle velocity field by evaluating the displacement per mesh vertex of the previous to the current time step and applying the interpolated velocities to the according grid cells of the obstacle velocity field.
Afterwards, the obstacle velocity field values are averaged to represent a correct discretization.

In \myrefalg{mov_smoke_sim} the simulation procedure of the moving smoke dataset is shown.
For our obstacle datasets the procedure in \myrefalg{rot_mov_cup_sim} is used,
with the prediction algorithm given in \myrefalg{rot_mov_cup_pred}.
Boundary conditions are abbreviated with BC in these algorithm.

\begin{algorithm}
   \caption{Moving smoke simulation}\label{alg:mov_smoke_sim}
   \begin{algorithmic}
      \WHILE{ $t \rightarrow t+1$ }
         \STATE $\rho \gets$ applyInflowSource($\rho$, $s$)
         \STATE $\rho \gets$ advect($\rho$, $\vu$)
         \STATE $\vu \gets$ advect($\vu$, $\vu$)
         \STATE $f \gets$ setWallBCs($f$, $\vu$) 
         \STATE $\vu \gets$ addBuoyancy($\rho$, $\vu$, $f$, $\vg$) 
         \STATE $p \gets$ solvePressure($f$, $\vu$) 
         \STATE $\vu \gets$ correctVelocity($\vu$, $p$) 
      \ENDWHILE
   \end{algorithmic}
\end{algorithm}

\begin{algorithm}
   \caption{Rotating and moving cup}\label{alg:rot_mov_cup_sim}
   \begin{algorithmic}[1]
      \WHILE{ $t \rightarrow t+1$ }
         \STATE $\rho \gets$ applyInflowSource($\rho$, $s$)
         \STATE $\rho \gets$ advect($\rho$, $\vu$)
         \STATE $\vu \gets$ advect($\vu$, $\vu$)
         \STATE $\vu_{obs} \gets$ computeObstacleVelocity($obstacle^{t}$, $obstacle^{t+1}$)
         \STATE $f \gets$ setObstacleFlags($obstacle^{t}$) 
         \STATE $f \gets$ setWallBCs($f$, $\vu$, $obstacle^{t}$, $\vu_{obs}$) 
         \STATE $\vu \gets$ addBuoyancy($\rho$, $\vu$, $f$, $\vg$) 
         \STATE $p \gets$ solvePressure($f$, $\vu$) 
         \STATE $\vu \gets$ correctVelocity($\vu$, $p$) 
      \ENDWHILE
   \end{algorithmic}
\end{algorithm}

\begin{algorithm}
   \caption{Rotating and moving cup network prediction \textit{Vel}}\label{alg:rot_mov_cup_pred}
   \begin{algorithmic}[1]
      \WHILE{ $t \rightarrow t+1$ }
         \STATE $\rho \gets$ applyInflowSource($\rho$, $s$)
         \STATE $\rho \gets$ advect($\rho$, $\vu$)
         \STATE $\vu \gets$ advect($\vu$, $\vu$)
         \STATE $\vu_{obs} \gets$ computeObstacleVelocity($obstacle^{t}$, $obstacle^{t+1}$)
         \STATE $f \gets$ setObstacleFlags($obstacle^{t}$) 
         \STATE $f \gets$ setWallBCs($f$, $\vu$, $obstacle^{t}$, $\vu_{obs}$) 

         \STATE $\dot{\vc}^{t} \gets$ encode($\tilde{\vu}^{t}$, $\rho^{t}$)
         \STATE $\hat{\vc}^{t} \gets [\tilde{\vc}^{t}_{vel}, \dot{\vc}^{t}_{den}]$
         \STATE $\tilde{\vc}^{t+1} \gets$ predict($\hat{\vc}^{t-1}$, $\hat{\vc}^{t}$)

         \STATE $\tilde{\vu}^{t+1}, \tilde{\rho}^{t+1} \gets$ decode($\tilde{\vc}^{t+1}$)\COMMENT $\tilde{\rho}^{t+1}$ is not used
         \STATE $\vu^{t+1} \gets \tilde{\vu}^{t+1}$ \COMMENT overwrite the velocity with the prediction
      \ENDWHILE
   \end{algorithmic}
\end{algorithm}

\subsection{Training Datasets}

In the following multiple simulations contained in our training data set are displayed.

\begin{figure}[ht]
   \newcommand\subfigwidth{.045\textwidth}
   \begin{center}
     \begin{subfigure}{\subfigwidth}
       \centering
       \includegraphics[width = 0.99\textwidth]{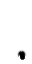}
     \end{subfigure}
     \begin{subfigure}{\subfigwidth}
       \centering
       \includegraphics[width = 0.99\textwidth]{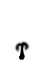}
     \end{subfigure}
      \begin{subfigure}{\subfigwidth}
         \centering
         \includegraphics[width = 0.99\textwidth]{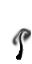}
       \end{subfigure}%
       \begin{subfigure}{\subfigwidth}
          \centering
          \includegraphics[width = 0.99\textwidth]{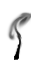}
       \end{subfigure}
       \begin{subfigure}{\subfigwidth}
         \centering
         \includegraphics[width = 0.99\textwidth]{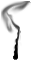}
       \end{subfigure}%
       \begin{subfigure}{\subfigwidth}
         \centering
         \includegraphics[width = 0.99\textwidth]{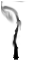}
       \end{subfigure}%
       \begin{subfigure}{\subfigwidth}
         \centering
         \includegraphics[width = 0.99\textwidth]{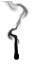}
       \end{subfigure}%
       \begin{subfigure}{\subfigwidth}
         \centering
         \includegraphics[width = 0.99\textwidth]{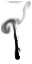}
       \end{subfigure}%
       \begin{subfigure}{\subfigwidth}
         \centering
         \includegraphics[width = 0.99\textwidth]{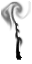}
       \end{subfigure}%
       \begin{subfigure}{\subfigwidth}
         \centering
         \includegraphics[width = 0.99\textwidth]{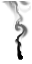}
       \end{subfigure}%
   \end{center}
   \begin{center}
     \begin{subfigure}{\subfigwidth}
       \centering
       \includegraphics[width = 0.99\textwidth]{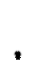}
    \end{subfigure}
     \begin{subfigure}{\subfigwidth}
        \centering
        \includegraphics[width = 0.99\textwidth]{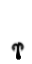}
     \end{subfigure}
     \begin{subfigure}{\subfigwidth}
        \centering
        \includegraphics[width = 0.99\textwidth]{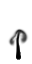}
      \end{subfigure}%
      \begin{subfigure}{\subfigwidth}
         \centering
         \includegraphics[width = 0.99\textwidth]{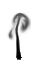}
      \end{subfigure}
      \begin{subfigure}{\subfigwidth}
        \centering
        \includegraphics[width = 0.99\textwidth]{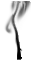}
      \end{subfigure}%
      \begin{subfigure}{\subfigwidth}
        \centering
        \includegraphics[width = 0.99\textwidth]{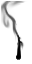}
      \end{subfigure}%
      \begin{subfigure}{\subfigwidth}
        \centering
        \includegraphics[width = 0.99\textwidth]{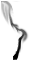}
      \end{subfigure}%
      \begin{subfigure}{\subfigwidth}
        \centering
        \includegraphics[width = 0.99\textwidth]{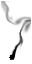}
      \end{subfigure}%
      \begin{subfigure}{\subfigwidth}
        \centering
        \includegraphics[width = 0.99\textwidth]{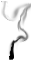}
      \end{subfigure}%
      \begin{subfigure}{\subfigwidth}
       \centering
       \includegraphics[width = 0.99\textwidth]{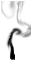}
     \end{subfigure}%
 \end{center}
 \begin{center}
   \begin{subfigure}{\subfigwidth}
     \centering
     \includegraphics[width = 0.99\textwidth]{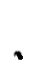}
   \end{subfigure}
   \begin{subfigure}{\subfigwidth}
      \centering
      \includegraphics[width = 0.99\textwidth]{images/dataset/smoke_mov/neg_000640.png}
   \end{subfigure}
   \begin{subfigure}{\subfigwidth}
      \centering
      \includegraphics[width = 0.99\textwidth]{images/dataset/smoke_mov/neg_000660.png}
    \end{subfigure}%
    \begin{subfigure}{\subfigwidth}
       \centering
       \includegraphics[width = 0.99\textwidth]{images/dataset/smoke_mov/neg_000680.png}
    \end{subfigure}
    \begin{subfigure}{\subfigwidth}
      \centering
      \includegraphics[width = 0.99\textwidth]{images/dataset/smoke_mov/neg_000700.png}
    \end{subfigure}%
    \begin{subfigure}{\subfigwidth}
      \centering
      \includegraphics[width = 0.99\textwidth]{images/dataset/smoke_mov/neg_000720.png}
    \end{subfigure}%
    \begin{subfigure}{\subfigwidth}
      \centering
      \includegraphics[width = 0.99\textwidth]{images/dataset/smoke_mov/neg_000740.png}
    \end{subfigure}%
    \begin{subfigure}{\subfigwidth}
      \centering
      \includegraphics[width = 0.99\textwidth]{images/dataset/smoke_mov/neg_000760.png}
    \end{subfigure}%
    \begin{subfigure}{\subfigwidth}
      \centering
      \includegraphics[width = 0.99\textwidth]{images/dataset/smoke_mov/neg_000780.png}
    \end{subfigure}%
    \begin{subfigure}{\subfigwidth}
     \centering
     \includegraphics[width = 0.99\textwidth]{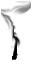}
   \end{subfigure}%
 \end{center}
 \begin{center}
   \begin{subfigure}{\subfigwidth}
     \centering
     \includegraphics[width = 0.99\textwidth]{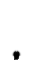}
   \end{subfigure}
   \begin{subfigure}{\subfigwidth}
      \centering
      \includegraphics[width = 0.99\textwidth]{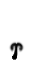}
   \end{subfigure}
   \begin{subfigure}{\subfigwidth}
      \centering
      \includegraphics[width = 0.99\textwidth]{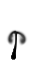}
    \end{subfigure}%
    \begin{subfigure}{\subfigwidth}
       \centering
       \includegraphics[width = 0.99\textwidth]{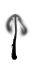}
    \end{subfigure}
    \begin{subfigure}{\subfigwidth}
      \centering
      \includegraphics[width = 0.99\textwidth]{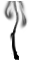}
    \end{subfigure}%
    \begin{subfigure}{\subfigwidth}
      \centering
      \includegraphics[width = 0.99\textwidth]{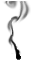}
    \end{subfigure}%
    \begin{subfigure}{\subfigwidth}
      \centering
      \includegraphics[width = 0.99\textwidth]{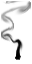}
    \end{subfigure}%
    \begin{subfigure}{\subfigwidth}
      \centering
      \includegraphics[width = 0.99\textwidth]{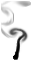}
    \end{subfigure}%
    \begin{subfigure}{\subfigwidth}
      \centering
      \includegraphics[width = 0.99\textwidth]{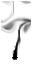}
    \end{subfigure}%
    \begin{subfigure}{\subfigwidth}
     \centering
     \includegraphics[width = 0.99\textwidth]{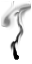}
   \end{subfigure}%
 \end{center}
   \caption{ Four example sequences of our moving smoke dataset. For visualization purposes we display frames $20$ to $200$ with a step size of $20$ for the respective scenes. The smoke density is shown as black.}
   \label{fig:ap_mov_smoke_data}
 \end{figure}
 
 \begin{figure}[ht]
   \newcommand\subfigwidth{.055\textwidth}
   \begin{center}
      \begin{subfigure}{\subfigwidth}
         \centering
         \includegraphics[width = 0.99\textwidth]{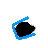}
      \end{subfigure}
      \begin{subfigure}{\subfigwidth}
         \centering
         \includegraphics[width = 0.99\textwidth]{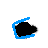}
       \end{subfigure}%
       \begin{subfigure}{\subfigwidth}
          \centering
          \includegraphics[width = 0.99\textwidth]{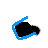}
       \end{subfigure}
       \begin{subfigure}{\subfigwidth}
         \centering
         \includegraphics[width = 0.99\textwidth]{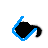}
       \end{subfigure}%
       \begin{subfigure}{\subfigwidth}
         \centering
         \includegraphics[width = 0.99\textwidth]{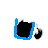}
       \end{subfigure}%
       \begin{subfigure}{\subfigwidth}
         \centering
         \includegraphics[width = 0.99\textwidth]{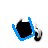}
       \end{subfigure}%
       \begin{subfigure}{\subfigwidth}
         \centering
         \includegraphics[width = 0.99\textwidth]{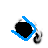}
       \end{subfigure}%
       \begin{subfigure}{\subfigwidth}
         \centering
         \includegraphics[width = 0.99\textwidth]{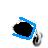}
       \end{subfigure}%
   \end{center}
   \begin{center}
     \begin{subfigure}{\subfigwidth}
        \centering
        \includegraphics[width = 0.99\textwidth]{images/dataset/smoke_rot_cup/composite_000240.png}
     \end{subfigure}
     \begin{subfigure}{\subfigwidth}
        \centering
        \includegraphics[width = 0.99\textwidth]{images/dataset/smoke_rot_cup/composite_000260.png}
      \end{subfigure}%
      \begin{subfigure}{\subfigwidth}
         \centering
         \includegraphics[width = 0.99\textwidth]{images/dataset/smoke_rot_cup/composite_000280.png}
      \end{subfigure}
      \begin{subfigure}{\subfigwidth}
        \centering
        \includegraphics[width = 0.99\textwidth]{images/dataset/smoke_rot_cup/composite_000300.png}
      \end{subfigure}%
      \begin{subfigure}{\subfigwidth}
        \centering
        \includegraphics[width = 0.99\textwidth]{images/dataset/smoke_rot_cup/composite_000320.png}
      \end{subfigure}%
      \begin{subfigure}{\subfigwidth}
        \centering
        \includegraphics[width = 0.99\textwidth]{images/dataset/smoke_rot_cup/composite_000340.png}
      \end{subfigure}%
      \begin{subfigure}{\subfigwidth}
        \centering
        \includegraphics[width = 0.99\textwidth]{images/dataset/smoke_rot_cup/composite_000360.png}
      \end{subfigure}%
      \begin{subfigure}{\subfigwidth}
        \centering
        \includegraphics[width = 0.99\textwidth]{images/dataset/smoke_rot_cup/composite_000380.png}
      \end{subfigure}%
   \end{center}
   \begin{center}
     \begin{subfigure}{\subfigwidth}
        \centering
        \includegraphics[width = 0.99\textwidth]{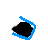}
     \end{subfigure}
     \begin{subfigure}{\subfigwidth}
        \centering
        \includegraphics[width = 0.99\textwidth]{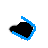}
      \end{subfigure}%
      \begin{subfigure}{\subfigwidth}
         \centering
         \includegraphics[width = 0.99\textwidth]{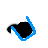}
      \end{subfigure}
      \begin{subfigure}{\subfigwidth}
        \centering
        \includegraphics[width = 0.99\textwidth]{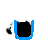}
      \end{subfigure}%
      \begin{subfigure}{\subfigwidth}
        \centering
        \includegraphics[width = 0.99\textwidth]{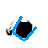}
      \end{subfigure}%
      \begin{subfigure}{\subfigwidth}
        \centering
        \includegraphics[width = 0.99\textwidth]{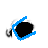}
      \end{subfigure}%
      \begin{subfigure}{\subfigwidth}
        \centering
        \includegraphics[width = 0.99\textwidth]{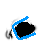}
      \end{subfigure}%
      \begin{subfigure}{\subfigwidth}
        \centering
        \includegraphics[width = 0.99\textwidth]{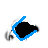}
      \end{subfigure}%
   \end{center}
   \caption{ Three example sequences of our rotating cup dataset. For visualization purposes we display frames $40$ to $180$ with a step size of $20$ for the respective scenes. The cup-shaped obstacle is highlighted in blue, whereas the smoke density is shown as black.}
   \label{fig:ap_rot_cup_data}
 \end{figure}

 \begin{figure}[ht]
   \newcommand\subfigwidth{.055\textwidth}
   \begin{center}
      \begin{subfigure}{\subfigwidth}
         \centering
         \frame{\includegraphics[width = 0.99\textwidth]{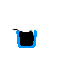}}
      \end{subfigure}%
      \begin{subfigure}{\subfigwidth}
         \centering
         \frame{\includegraphics[width = 0.99\textwidth]{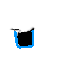}}
       \end{subfigure}%
       \begin{subfigure}{\subfigwidth}
         \centering
         \frame{\includegraphics[width = 0.99\textwidth]{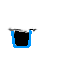}}
       \end{subfigure}%
       \begin{subfigure}{\subfigwidth}
         \centering
         \frame{\includegraphics[width = 0.99\textwidth]{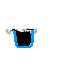}}
       \end{subfigure}%
       \begin{subfigure}{\subfigwidth}
         \centering
         \frame{\includegraphics[width = 0.99\textwidth]{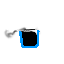}}
       \end{subfigure}%
       \begin{subfigure}{\subfigwidth}
         \centering
         \frame{\includegraphics[width = 0.99\textwidth]{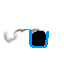}}
       \end{subfigure}%
       \begin{subfigure}{\subfigwidth}
         \centering
         \frame{\includegraphics[width = 0.99\textwidth]{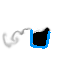}}
       \end{subfigure}%
       \begin{subfigure}{\subfigwidth}
         \centering
         \frame{\includegraphics[width = 0.99\textwidth]{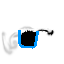}}
       \end{subfigure}%
   \end{center}
   \begin{center}
     \begin{subfigure}{\subfigwidth}
        \centering
        \frame{\includegraphics[width = 0.99\textwidth]{images/dataset/smoke_rot_cup_mov/composite_000640.png}}
     \end{subfigure}%
     \begin{subfigure}{\subfigwidth}
        \centering
        \frame{\includegraphics[width = 0.99\textwidth]{images/dataset/smoke_rot_cup_mov/composite_000660.png}}
      \end{subfigure}%
      \begin{subfigure}{\subfigwidth}
         \centering
         \frame{\includegraphics[width = 0.99\textwidth]{images/dataset/smoke_rot_cup_mov/composite_000680.png}}
      \end{subfigure}%
      \begin{subfigure}{\subfigwidth}
        \centering
        \frame{\includegraphics[width = 0.99\textwidth]{images/dataset/smoke_rot_cup_mov/composite_000700.png}}
      \end{subfigure}%
      \begin{subfigure}{\subfigwidth}
        \centering
        \frame{\includegraphics[width = 0.99\textwidth]{images/dataset/smoke_rot_cup_mov/composite_000720.png}}
      \end{subfigure}%
      \begin{subfigure}{\subfigwidth}
        \centering
        \frame{\includegraphics[width = 0.99\textwidth]{images/dataset/smoke_rot_cup_mov/composite_000740.png}}
      \end{subfigure}%
      \begin{subfigure}{\subfigwidth}
        \centering
        \frame{\includegraphics[width = 0.99\textwidth]{images/dataset/smoke_rot_cup_mov/composite_000760.png}}
      \end{subfigure}%
      \begin{subfigure}{\subfigwidth}
        \centering
        \frame{\includegraphics[width = 0.99\textwidth]{images/dataset/smoke_rot_cup_mov/composite_000780.png}}
      \end{subfigure}%
   \end{center}
   \begin{center}
     \begin{subfigure}{\subfigwidth}
        \centering
        \frame{\includegraphics[width = 0.99\textwidth]{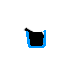}}
     \end{subfigure}%
     \begin{subfigure}{\subfigwidth}
        \centering
        \frame{\includegraphics[width = 0.99\textwidth]{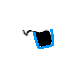}}
      \end{subfigure}%
      \begin{subfigure}{\subfigwidth}
         \centering
         \frame{\includegraphics[width = 0.99\textwidth]{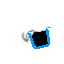}}
      \end{subfigure}%
      \begin{subfigure}{\subfigwidth}
        \centering
        \frame{\includegraphics[width = 0.99\textwidth]{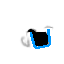}}
      \end{subfigure}%
      \begin{subfigure}{\subfigwidth}
        \centering
        \frame{\includegraphics[width = 0.99\textwidth]{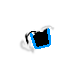}}
      \end{subfigure}%
      \begin{subfigure}{\subfigwidth}
        \centering
        \frame{\includegraphics[width = 0.99\textwidth]{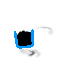}}
      \end{subfigure}%
      \begin{subfigure}{\subfigwidth}
        \centering
        \frame{\includegraphics[width = 0.99\textwidth]{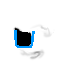}}
      \end{subfigure}%
      \begin{subfigure}{\subfigwidth}
        \centering
        \frame{\includegraphics[width = 0.99\textwidth]{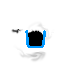}}
      \end{subfigure}%
   \end{center}
   \caption{
   Three example sequences of our rotating and moving cup dataset. For visualization purposes we display frames $40$ to $180$ with a step size of $20$ for the respective scenes. The cup-shaped obstacle is highlighted in blue, whereas the smoke density is shown as black.}
   \label{fig:ap_rot_cup_mov_data}
 \end{figure}





\end{document}